\pdfoutput=1

\documentclass[11pt,twoside,a4paper,cmspaper,final,collab]{cms-tdr}
\def\svnVersion{485686P}\def\cmsCernNoTag{CERN-EP-2018-211}\def\cmsCernDate{\today}\def\cmsMessage{Published in Physical Review D as \href{http://dx.doi.org/10.1103/PhysRevD.98.112014}{\doi{10.1103/PhysRevD.98.112014.}}}

\begin{document}\cmsNoteHeader{EXO-17-021}

\hyphenation{had-ron-i-za-tion}
\hyphenation{cal-or-i-me-ter}
\hyphenation{de-vices}
\RCS$HeadURL: svn+ssh://svn.cern.ch/reps/tdr2/papers/EXO-17-021/trunk/EXO-17-021.tex $
\RCS$Id: EXO-17-021.tex 478986 2018-10-23 15:36:22Z alverson $

\newlength\cmsFigWidthTwo
\ifthenelse{\boolean{cms@external}}{\setlength\cmsFigWidthTwo{0.98\columnwidth}}{\setlength\cmsFigWidthTwo{0.49\textwidth}}
\ifthenelse{\boolean{cms@external}}{\providecommand{\cmsLeft}{top\xspace}}{\providecommand{\cmsLeft}{left\xspace}}
\ifthenelse{\boolean{cms@external}}{\providecommand{\cmsRight}{bottom\xspace}}{\providecommand{\cmsRight}{right\xspace}}
\ifthenelse{\boolean{cms@external}}{\providecommand{\NA}{\ensuremath{\cdots}\xspace}}{\providecommand{\NA}{\ensuremath{\text{---}}\xspace}}
\ifthenelse{\boolean{cms@external}}{\providecommand{\CL}{C.L.\xspace}}{\providecommand{\CL}{CL\xspace}}

\newcommand{\resHT}{\ensuremath{\HT^{\mathrm{AK4}}}\xspace}
\newcommand{\boosHT}{\ensuremath{\HT^{\mathrm{AK8}}}\xspace}
\newcommand{\qq}{\ensuremath{{\PQq}{\PQq}^{\prime}}\xspace}
\newcommand{\qpq}{\ensuremath{\PQq^{\prime}\PAQq}\xspace}
\newcommand{\bq}{\ensuremath{\cPqb\cPq^{\prime}}\xspace}
\newcommand{\stopqq}{\ensuremath{\PSQt\to\cPq\cPq^{\prime}}\xspace}
\newcommand{\stopbq}{\ensuremath{\PSQt\to\cPqb\cPq^{\prime}}\xspace}
\newcommand{\UDDqq}{\ensuremath{\lambda^{\prime \prime}_{{312}}}\xspace}
\newcommand{\UDDbq}{\ensuremath{\lambda^{\prime \prime}_{{323}}}\xspace}
\newcommand{\threeDD}{\ensuremath{\lambda^{\prime \prime}_{\mathrm{3DD}}}\xspace}
\newcommand{\ijk}{\ensuremath{\lambda^{\prime \prime}_{ijk}}\xspace}
\newcommand{\stopmass}{\ensuremath{m_{\PSQt}}\xspace}
\newcommand{\prunedAveMass}{\ensuremath{\overline{m}}\xspace}
\newcommand{\masym}{\ensuremath{m_{\text{asym}}}\xspace}
\newcommand{\aveMass}{\ensuremath{\overline{M}}\xspace}
\newcommand{\Masym}{\ensuremath{M_{\text{asym}}}\xspace}
\newcommand{\deta}{\ensuremath{\Delta \eta}\xspace}
\newcommand{\Deta}{\ensuremath{\Delta \eta_{\text{dijet}}}\xspace}
\newcommand{\DeltaR}{\ensuremath{\Delta R_{\text{dijet}}}\xspace}
\newcommand{\tauN}[1]{\ensuremath{\tau_{#1}}\xspace}
\newcommand{\mj}[1]{\ensuremath{m_{\mathrm{j#1}}}\xspace}
\newcommand{\Mjj}[1]{\ensuremath{m_{\mathrm{jj#1}}}\xspace}
\newlength\cmsTabSkip\setlength{\cmsTabSkip}{1ex}

\cmsNoteHeader{EXO-17-021}
\title{Search for pair-produced resonances decaying to quark pairs in proton-proton collisions at \texorpdfstring{$\sqrt{s}=13$\TeV}{sqrt(s) = 13 TeV}}

\date{\today}

\abstract{
A general search for the pair production of resonances, each decaying to two quarks, is reported. The search is conducted separately for heavier resonances (masses above 400\GeV), where each of the four final-state quarks generates a hadronic jet resulting in a four-jet signature, and for lighter resonances (masses between 80 and 400\GeV), where the pair of quarks from each resonance is collimated and reconstructed as a single jet resulting in a two-jet signature. In addition, a {\cPqb}-tagged selection is applied to target resonances with a bottom quark in the final state. The analysis uses data collected with the CMS detector at the CERN LHC, corresponding to an integrated luminosity of 35.9\fbinv, from proton-proton collisions at a center-of-mass energy of 13\TeV. The mass spectra are analyzed for the presence of new resonances, and are found to be consistent with standard model expectations. The results are interpreted in the framework of $R$-parity-violating supersymmetry assuming the pair production of scalar top quarks decaying via the hadronic coupling \UDDqq or \UDDbq, and upper limits on the cross section as a function of the top squark mass are set. These results probe a wider range of masses than previously explored at the LHC, and extend the top squark mass limits in the \stopqq scenario.
}

\hypersetup{
pdfauthor={CMS Collaboration},
pdftitle={Search for pair-produced resonances decaying to quark pairs in proton-proton collisions at sqrt(s)= 13 TeV},
pdfsubject={CMS},
pdfkeywords={CMS, physics, dijets}}

\maketitle
\newpage

\section{Introduction}

New particles that decay into quarks and gluons and produce fully hadronic signatures are predicted in many models of physics beyond the standard model (SM)~\cite{Kilic:2008pm,Hill:1991at,Kribs:2007ac}.
For instance, the violation of baryon number in certain supersymmetric (SUSY) models leads to colored superpartners producing fully hadronic final states~\cite{Evans2012}.
In this paper, we report on a generic search for pair-produced resonances decaying to two light quarks (\qq) or one light quark and one bottom quark (\bq).

Minimal SUSY models introduce $R$-parity, associated with a $\Z_2$ symmetry group called $R$ symmetry, to forbid terms in the SUSY potential that naturally lead to the violation of baryon or lepton numbers~\cite{Barbier:2004ez}.
After SUSY breaking, $R$-parity violating Yukawa interactions of the form
\begin{linenomath}
\begin{equation}
\lambda_{ijk} L_{i} L_{j} E^{c}_{k} , \quad \lambda^{\prime}_{ijk} L_{i} Q_{j} D^{c}_{k} , \quad \ijk U^{c}_{i} D^{c}_{j} D^{c}_{k} ,\label{eq:RPV}
\end{equation}
\end{linenomath}
can appear in the Lagrangian, where $\lambda$, $\lambda^{\prime}$, $\lambda^{\prime \prime}$ are coupling constants, and $i,j,k$ are quark and lepton generation indices following the summation convention, while $c$ denotes charge conjugation. The $SU(2)$ doublet superfields of the lepton and quark are denoted by  $L_{i}$ and $Q_{i}$, respectively, while the $E_{i}$, $U_{i}$ and, $D_{j}$ represent the $SU(2)$ singlet superfields of the lepton, up- and down-type quarks, respectively. The first and third terms in Eq.~\eqref{eq:RPV} are antisymmetric in \{$i,j$\} and \{$j,k$\}, respectively.
The trilinear couplings \ijk permit vertices of sfermions interacting with two fermions, and in baryonic $R$-parity-violating (RPV) models, the only nonzero couplings in Eq.~\eqref{eq:RPV} are \ijk, which produce interactions of squarks with two quarks.

\begin{figure}[hbtp]
\centering
\includegraphics[width=.3\textwidth]{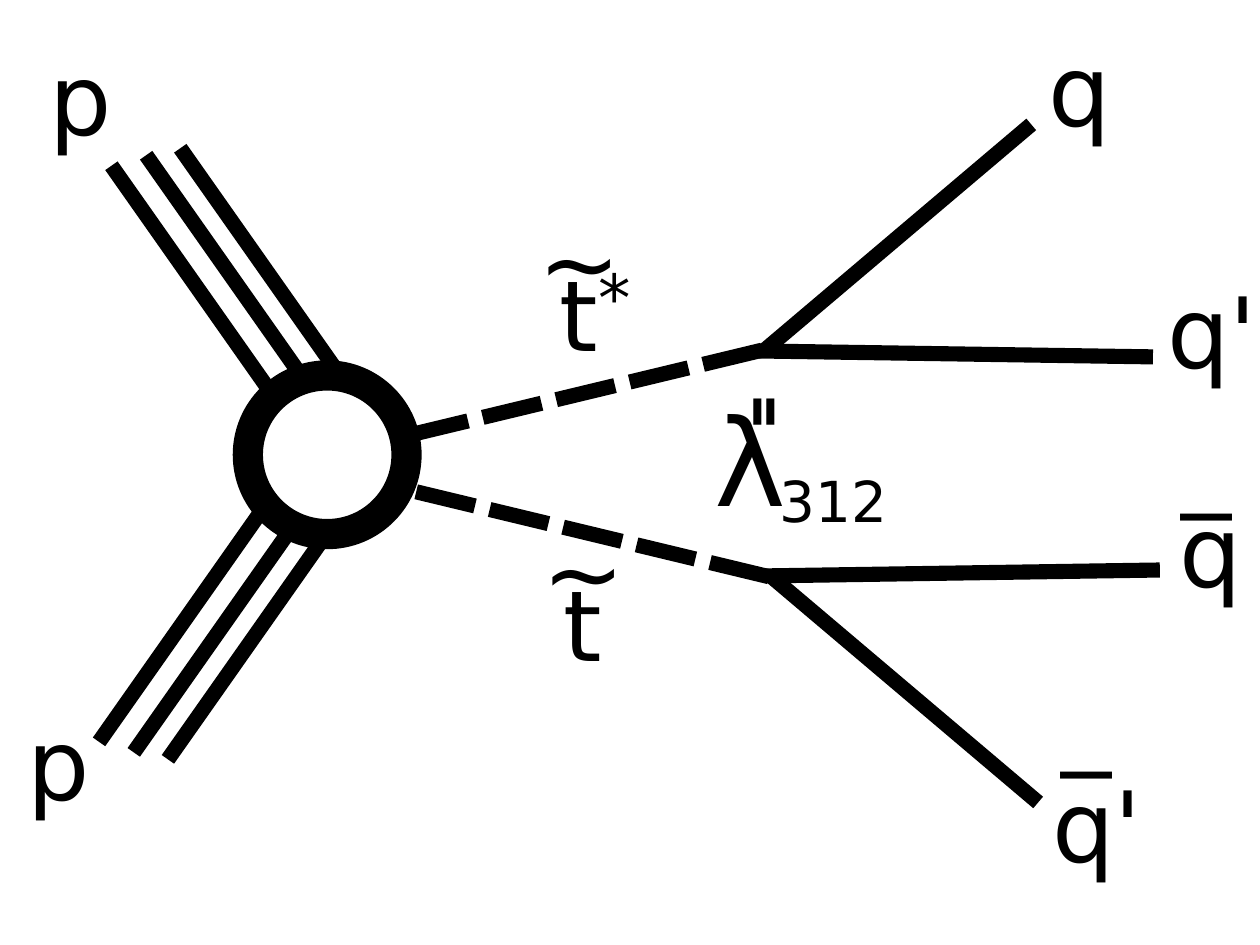}
\includegraphics[width=.3\textwidth]{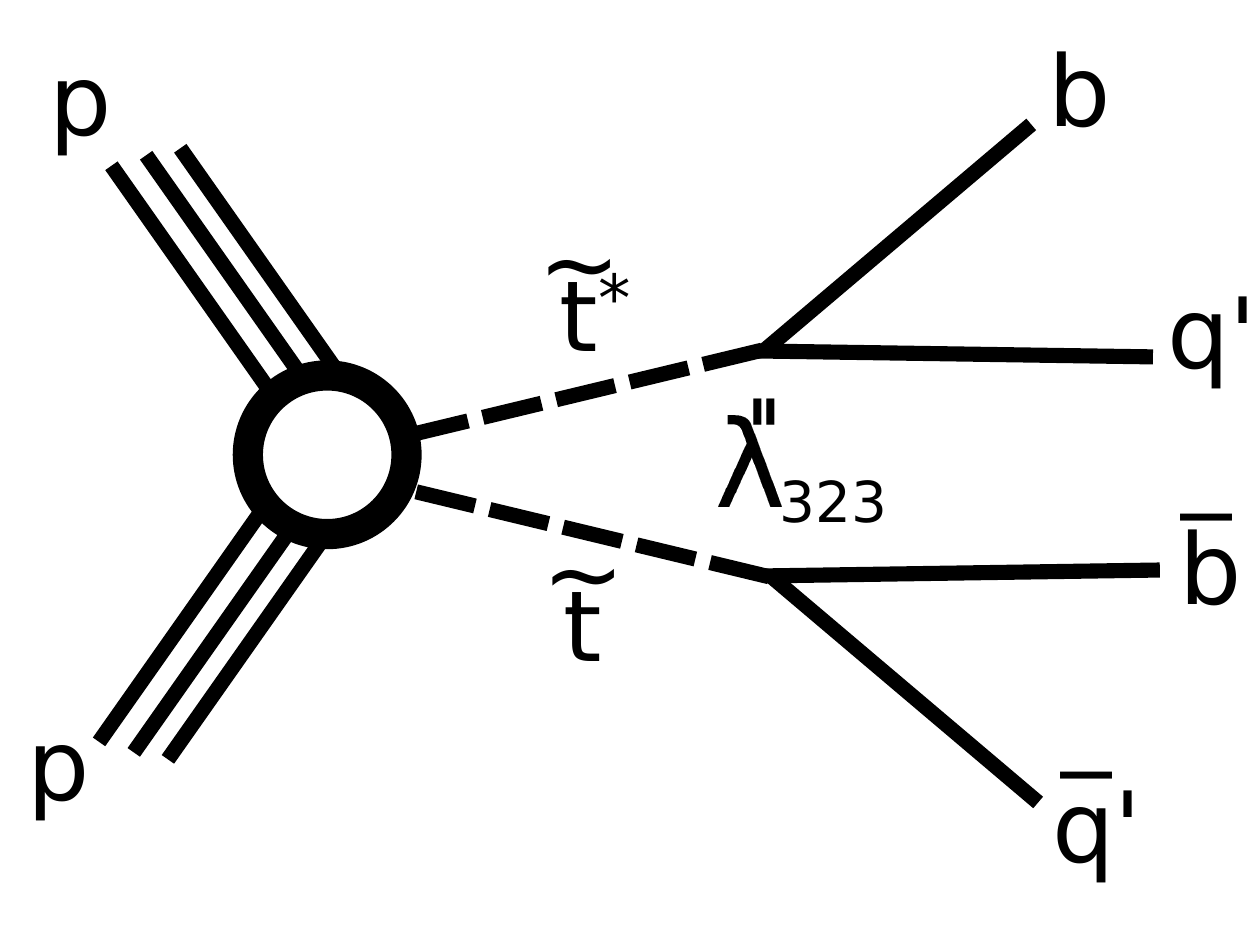}
	\caption{Diagrams for the benchmark models used in this analysis: pair production of top squarks decaying into \qq via the RPV coupling \UDDqq (\cmsLeft), and \bq via the RPV coupling \UDDbq (\cmsRight).}
\label{fig:feynman}
\end{figure}

We consider pair production of top squarks (\PSQt) as a benchmark model, assuming the \PSQt is the lightest of the colored SUSY partners and is allowed to decay via the baryonic RPV coupling to quarks.
In this case $\ijk = \threeDD$ and each index reflects the squark or quark generation of the process, two of which are down-type quarks.
Two possible choices of hadronic RPV coupling scenarios are studied: \stopqq through the coupling \UDDqq, and \stopbq through the coupling \UDDbq.
The couplings considered are assumed to be large enough such that the resulting decays are prompt.
These two models are schematically depicted in Fig.~\ref{fig:feynman}.

Searches for \stopqq via RPV decays have been performed at CERN by the ALEPH experiment at LEP~\cite{ALEPHPaper}, which excluded $\stopmass < 80\GeV$ at 95\% confidence level (\CL), and subsequently by the CDF experiment~\cite{CDFPaper} at the Fermilab Tevatron, which extended the limit to $\stopmass < 100 \GeV$.
Similar searches have been performed at the CERN LHC by both the CMS and ATLAS experiments at center-of-mass energies $\sqrt{s}=7$, 8, and 13\TeV; CMS~\cite{CMSRunIPaper} excluded $200 < \stopmass < 350\GeV$ at $\sqrt{s}=8\TeV$, while the ATLAS exclusion~\cite{ATLASLimits13TeV} is $100 < \stopmass < 410\GeV$ at $\sqrt{s}=13\TeV$.
For the \stopbq scenario, mass exclusion limits at $\sqrt{s}=8\TeV$ have been reported by CMS~\cite{CMSRunIPaper} of $200 < \stopmass < 385\GeV$, and by ATLAS~\cite{ATLASLimits8TeV} of $100 < \stopmass < 310\GeV$, and at $\sqrt{s}=13\TeV$ ATLAS~\cite{ATLASLimits13TeV} excluded $100 < \stopmass < 470\GeV$ and $480 < \stopmass < 610\GeV$.

{\tolerance=800 The analysis reported in this paper uses {\Pp\Pp} collision data at $\sqrt{s} = 13\TeV$ collected with the CMS detector~\cite{Chatrchyan:2008aa} at the LHC in 2016, corresponding to an integrated luminosity of 35.9\fbinv~\cite{CMS-PAS-LUM-17-001}.
The search is conducted in two mass ranges.
The mass spectrum between 60 and 450\GeV is used to search for lighter resonances between 80 and 400\GeV, where the decay products of each of the resonances are sufficiently collimated to be reconstructed as a single jet (boosted search).
The mass spectrum above 350\GeV is explored for the presence of heavier resonances above 400\GeV, where four jets are reconstructed in the final state (resolved search).
Together they target resonance masses between 80 and 1500\GeV.
When {\cPqb} tagging requirements are applied to either of the searches, we refer to the selection as {\cPqb} tagged, and interpret the results in the \stopbq scenario. When no {\cPqb} tagging is applied, we refer to the selection as inclusive, and interpret the results in the \stopqq scenario. In both searches, the selection criteria and analysis strategies are general, such that any pair produced diquark resonance with a narrow width and sufficient cross section would appear as a local enhancement in the mass spectra. \par}

The low-mass boosted search exploits the internal structure of the jets to differentiate between signal jets (two-prong structure) and jets coming from quantum chromodynamics (QCD) multijet processes (predominantly with no internal structure).
In this search, we use the average mass of the two jets with the highest transverse momentum (\pt), after removing soft and wide-angle QCD multijet radiation, to look for evidence of a signal consistent with localized deviations from the estimated SM backgrounds.
The primary SM background component---QCD multijet events---is estimated from data control samples.
Subdominant SM processes, such as the single and double production of {\PW} and {\cPZ} bosons, and top quarks decaying hadronically, are taken into account with simulated samples.
These backgrounds create resonances in the mass spectrum, and they are henceforth referred to as resonant backgrounds.

For the resolved search, the high-mass resonances are produced with insufficient boost for the decay products to be merged into single jets, and events with four individual high transverse momentum (\pt) jets are selected.
The dijet mass spectrum in this search is also dominated by QCD multijet production.
The mass spectrum is parameterized as a steeply falling smooth distribution that is explored for signal-like localized excesses.

\section{The CMS detector}

The central feature of the CMS apparatus is a superconducting solenoid of 6\unit{m} internal diameter, providing a magnetic field of 3.8\unit{T}.
Within the solenoid volume are a silicon pixel and a strip tracker, a lead tungstate crystal electromagnetic calorimeter (ECAL), and a brass and scintillator hadron calorimeter (HCAL), each composed of a barrel and two endcap sections.
Forward calorimeters extend the pseudorapidity ($\eta$) coverage provided by the barrel and endcap detectors.
Muons are detected in gas-ionization chambers embedded in the steel flux-return yoke outside the solenoid.
Energy deposits from hadronic jets are measured using the ECAL and HCAL.
Events of interest are selected using a two-tiered trigger system~\cite{Khachatryan:2016bia}.
A detailed description of the CMS detector, together with a definition of the coordinate system used and the relevant kinematic variables, can be found in Ref.~\cite{Chatrchyan:2008aa}.

\section{Simulated event samples}
\label{sec:dataset}

Top squark signal events are simulated using a combination of {\PYTHIA} 8.212~\cite{Sjostrand:2014zea} and {\MGvATNLO} 2.2.2~\cite{madgraph}.
The calculation of the production of a pair of top squarks with up to two additional initial-state radiation jets is performed at leading order (LO) with {\MGvATNLO} and MLM merging ~\cite{mlmmatch}, while \PYTHIA is used for the prompt decay of each top squark to either \stopqq or \stopbq through the \threeDD hadronic RPV couplings.
For each of the coupling models considered, all other $\lambda^{\prime\prime}_{\mathrm{UDD}}$ couplings are set to zero so that the branching fraction to the desired channel is 100\%.
The \PYTHIA simulation is also used for the parton showering and the fragmentation with the CUETP8M1~\cite{pythiaTune} underlying event tune.
For each coupling, top squarks are generated with masses between 80 and 1500\GeV, in 20\GeV increments up to 300\GeV, in 50\GeV steps up to 1\TeV, and in 100\GeV increments thereafter.
All other SUSY particle masses are set to higher values in order not to produce intermediate sparticles in the top squark production and decay.
The natural width of the top squark is taken to be much smaller than the detector resolution.

Processes from QCD multijets are simulated at LO via \PYTHIA using the CUETP8M1 tune~\cite{pythiaTune}.
The production of a hadronically decaying {\PW} or {\PZ} boson accompanied by additional jets from initial- and final-state radiation ($\PW{\to}\qpq{+}\text{jets}$ or $\cPZ{\to}\qqbar{+}\text{jets}$)~\cite{mlmmatch},
and {\cPZ}{\cPZ} diboson ~\cite{fxfxmatch} samples are generated with {\MGvATNLO}, at LO with MLM merging and at next-to-leading order (NLO) with FxFx merging~\cite{fxfxmatch}, respectively.
{\PW}{\cPZ} processes are generated at LO with \PYTHIA, and {\ttbar}+jets and {\PW}{\PW} samples are generated at NLO with {\POWHEG} v2~\cite{Alioli:2011as,Melia:2011gk}.
For $\PW{\to}\qpq{/}\cPZ{\to}\qqbar{+}\text{jets}$ events, higher-order \pt-dependent electroweak NLO corrections are applied to improve the modeling of the kinematic distributions~\cite{Kallweit:2015dum,Kallweit:2015fta,Lindert:2017olm,Sirunyan:2017hci,Kallweit:2014xda}.

Additional {\Pp\Pp} interactions in the same or adjacent bunch crossings are referred to as pileup.
A number of minimum bias interactions are added to the hard interaction of all simulated samples, and the events are weighted such that the distribution of the number of pileup interactions is the same as that in the data.
{\PYTHIA} is used for the parton showering and hadronization and the simulation of the CMS detector for all samples is handled by \GEANTfour~\cite{geant4}.
All simulated samples are produced with the parton distribution functions (PDF) NNPDF3.0~\cite{Ball:2014uwa}, with the precision (LO or NLO) set by the generator used.

\section{Jet reconstruction and selection}
\label{sec:object}

The reconstructed vertex with the largest value of summed physics-object $\pt^2$ is taken to be the primary $\Pp\Pp$ interaction vertex.
Here the physics objects are the jets, clustered using the anti-\kt jet finding algorithm~\cite{Cacciari:2008gp,Cacciari:2011ma}, with the tracks assigned to the vertex as inputs, and the associated missing transverse momentum, taken as the negative vector \pt sum of those jets.
Particle candidates in CMS are reconstructed using a particle-flow (PF) algorithm~\cite{pflow}, which identifies muons, electrons, photons, and neutral and charged hadrons through a combination of information from the various subdetectors.
The PF candidates identified as originating from pileup are removed prior to the jet clustering~\cite{Cacciari:2007fd,Khachatryan:2016kdb}.
Jets with a clustering distance parameter of 0.4 (AK4 jets) and 0.8 (AK8 jets) are used for the resolved and the boosted searches, respectively.
Corrections are applied to jet energies as a function of $\eta$ and \pt of the jet to account for the combined response function of the detector to reconstructed objects~\cite{1748-0221-6-11-P11002,Khachatryan:2016kdb}.

For the boosted search, jet grooming techniques are used to eliminate soft, and wide-angle QCD radiation at the periphery of the jet.
Grooming improves the jet mass resolution and reduces the pileup contributions to the jet mass.
Two grooming algorithms are used: trimming~\cite{Trimming} at the trigger stage and pruning~\cite{Pruning} at the analysis stage.
The trimming technique discriminates particles within the constituents of the jet based on a dynamic \pt threshold.
In pruning, the constituents of the original jet are reclustered with the same distant parameter but using a modified Cambridge--Aachen (CA) algorithm~\cite{CAref1,CAref2} with relative \pt and angular requirements.
To discriminate between jets originating from SM background processes from those from boosted hadronic resonances, $N$-subjetiness variables (\tauN{N})~\cite{nsubjetiness} are used,
which quantify the number of $N$ prongs of energy inside a jet.   In particular, ratios of $N$-subjetiness variables, $\tauN{MN} = \tauN{M}/\tauN{N}$, are found to provide better discrimination between signal and background.
In this analysis,   $\tauN{21} = \tauN{2}/ \tauN{1}$ is used to distinguish two-prong signal-like jets and one-prong background-like jets which arise from QCD multijets events at an overwhelming rate, and $\tauN{32} = \tauN{3}/ \tauN{2}$ to separate two-prong jets from three-prong jets from hadronically decaying top quarks.

Jets produced by the hadronization of bottom quarks are identified with a combined secondary vertex {\cPqb}-tagging algorithm~\cite{BTV-16-002}.
This algorithm uses a multivariate discriminator with inputs from information related to the secondary vertex, and a track-based lifetime measurement to differentiate between jets from bottom quarks and from light-flavor quarks and gluons.
The working point of the {\cPqb}-tagging algorithm used in this analysis is referred to as loose, and gives an ${\approx}81\%$ {\cPqb} tagging efficiency, a ${\approx}10\%$ misidentification rate for light-quark and gluon jets, and a ${\approx}40\%$ misidentification rate for charm quark jets~\cite{BTV-16-002}.

\section{Boosted search}
\label{sec:boosted}

\subsection{Event selection}
Events are first selected with a trigger based on the total hadronic transverse momentum in the event (\HT), defined as the scalar \pt sum of AK4 jets (\resHT) with $\pt>30\GeV$ and $\abs{\eta}< 2.5$.
The \resHT trigger threshold for the early data-taking period was set to 800\GeV, and raised to 900\GeV for the last 8\fbinv of data to enable the trigger to handle the instantaneous luminosity delivered by the LHC.
Additionally, we include a logical OR of two triggers based on AK8 jets: one trigger requires an AK8 jet with $\pt>360\GeV$ and trimmed mass above 30\GeV, the other requires $\boosHT>750\GeV$ defined with AK8 jets with $\pt>150\GeV$, and a jet with trimmed mass above 50\GeV.
The selection efficiency of the chosen triggers is determined relative to unbiased samples collected with muon based triggers. This is cross checked with other samples collected with jet based triggers, and are all found to give consistent results.
The signal triggers are found to have an efficiency greater than 98\% with respect to the analysis-level selection, for events satisfying $\boosHT>900\GeV$.
In addition to satisfying the trigger conditions, selected events are required to have at least two AK8 jets with $\pt> 150\GeV$, situated in the central region of the detector with $\abs{\eta}<2.5$, and $\boosHT>900\GeV$.

The boosted search assumes that the decay products of the resonance would be fully contained in a very energetic AK8 jet, and therefore we select the two most energetic AK8 jets in the event.
The pruning algorithm is used to compute the mass of each of these two jets (\mj{1} and \mj{2}).
The spectrum of the average pruned jet mass of these two jets, $\prunedAveMass = (m_{\mathrm{j1}}+m_{\mathrm{j2}})/2$, is examined for the presence of new physics in the mass range 60--450\GeV.

\begin{figure*}
\centering
	\includegraphics[width=\cmsFigWidthTwo]{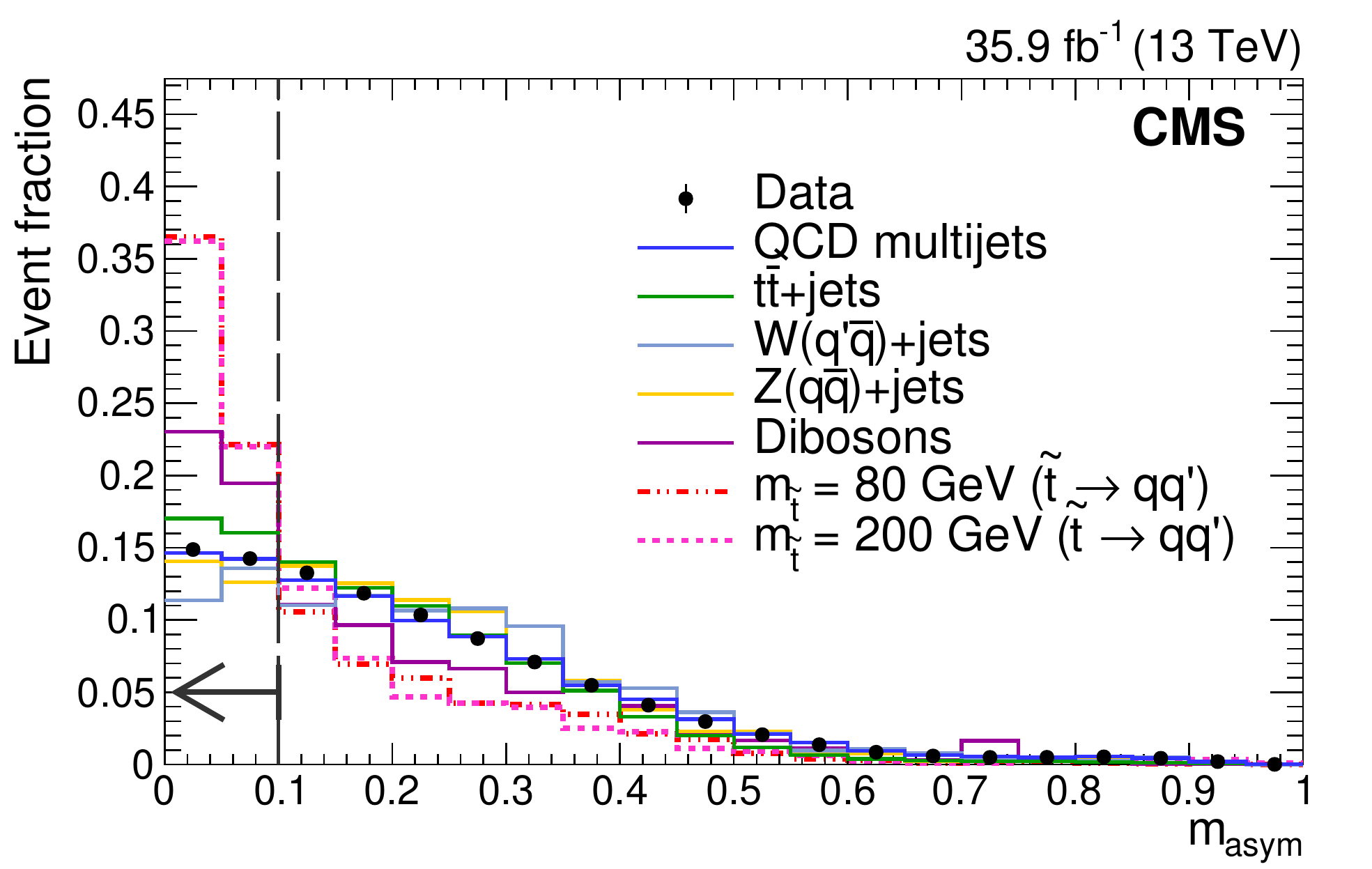}
	\includegraphics[width=\cmsFigWidthTwo]{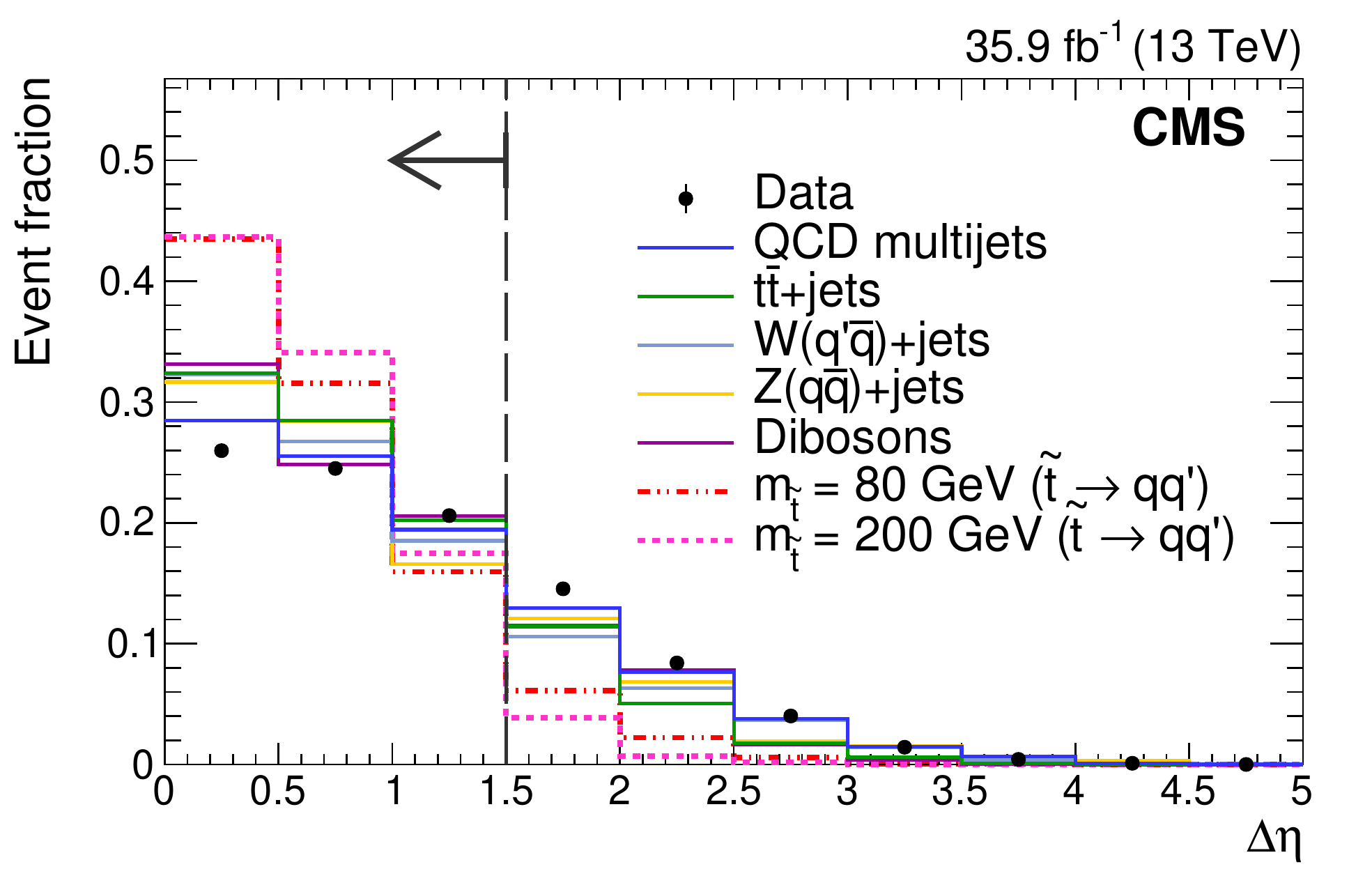}
	\includegraphics[width=\cmsFigWidthTwo]{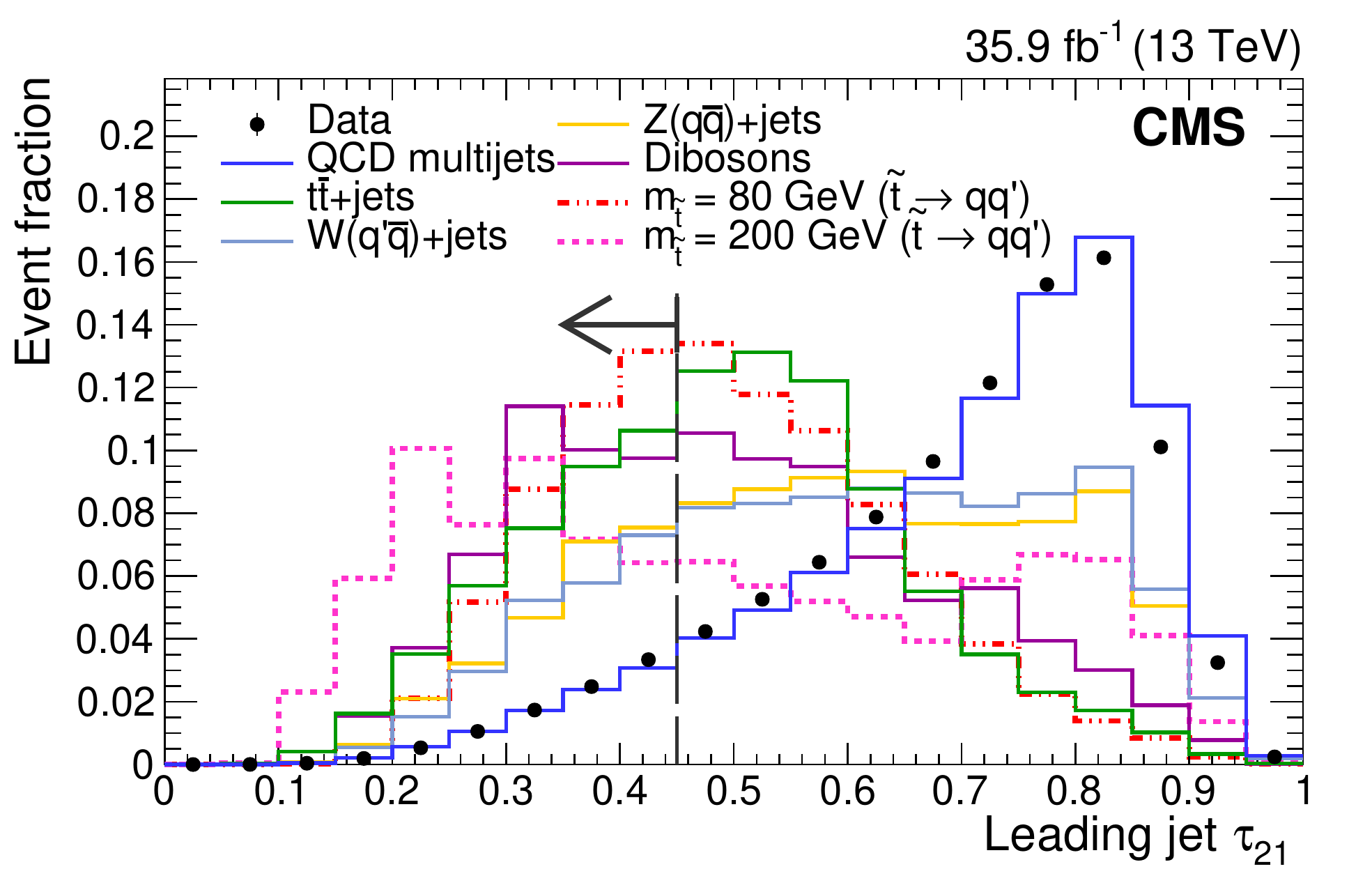}
	\includegraphics[width=\cmsFigWidthTwo]{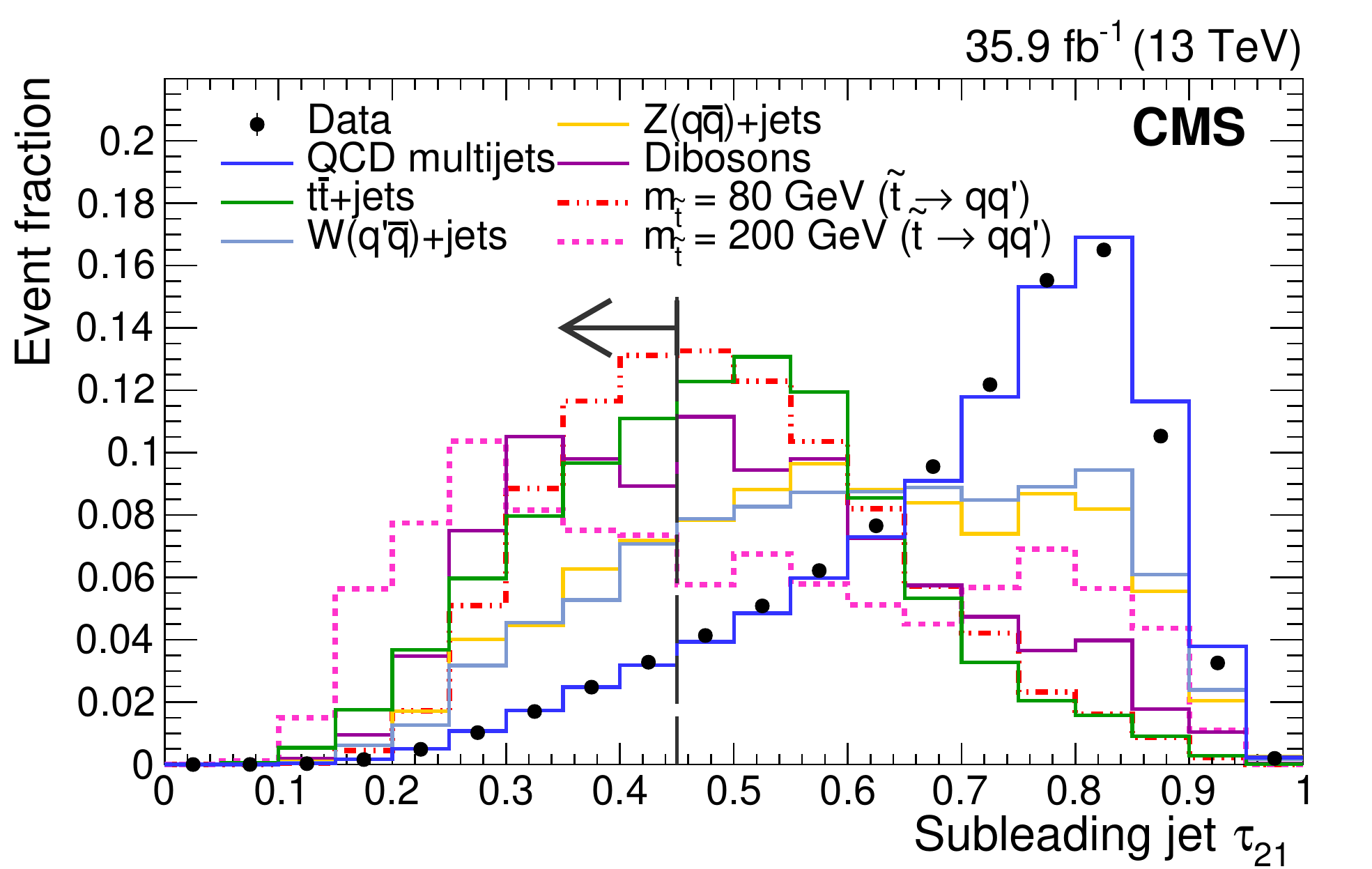}
	\includegraphics[width=\cmsFigWidthTwo]{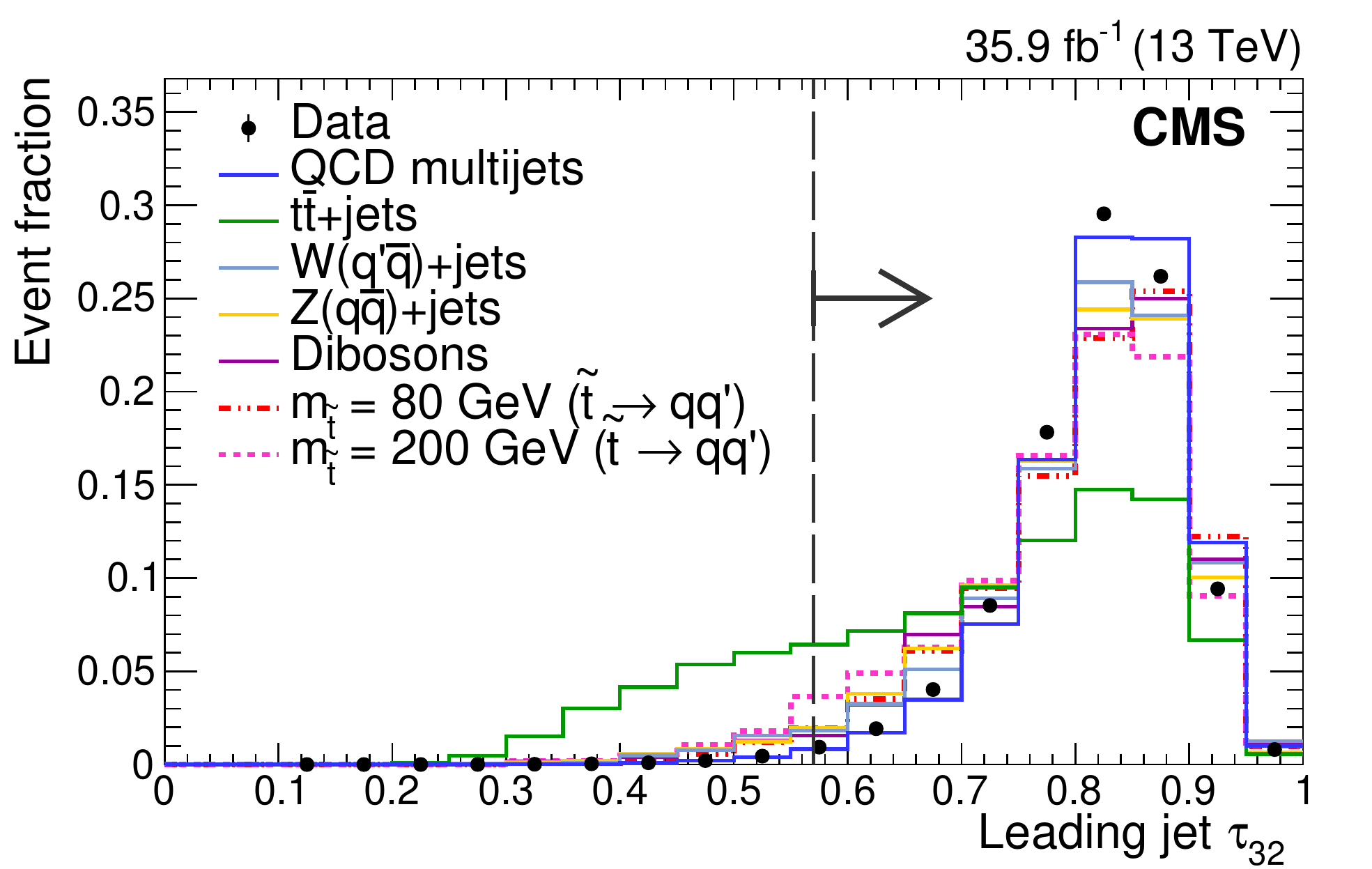}
	\includegraphics[width=\cmsFigWidthTwo]{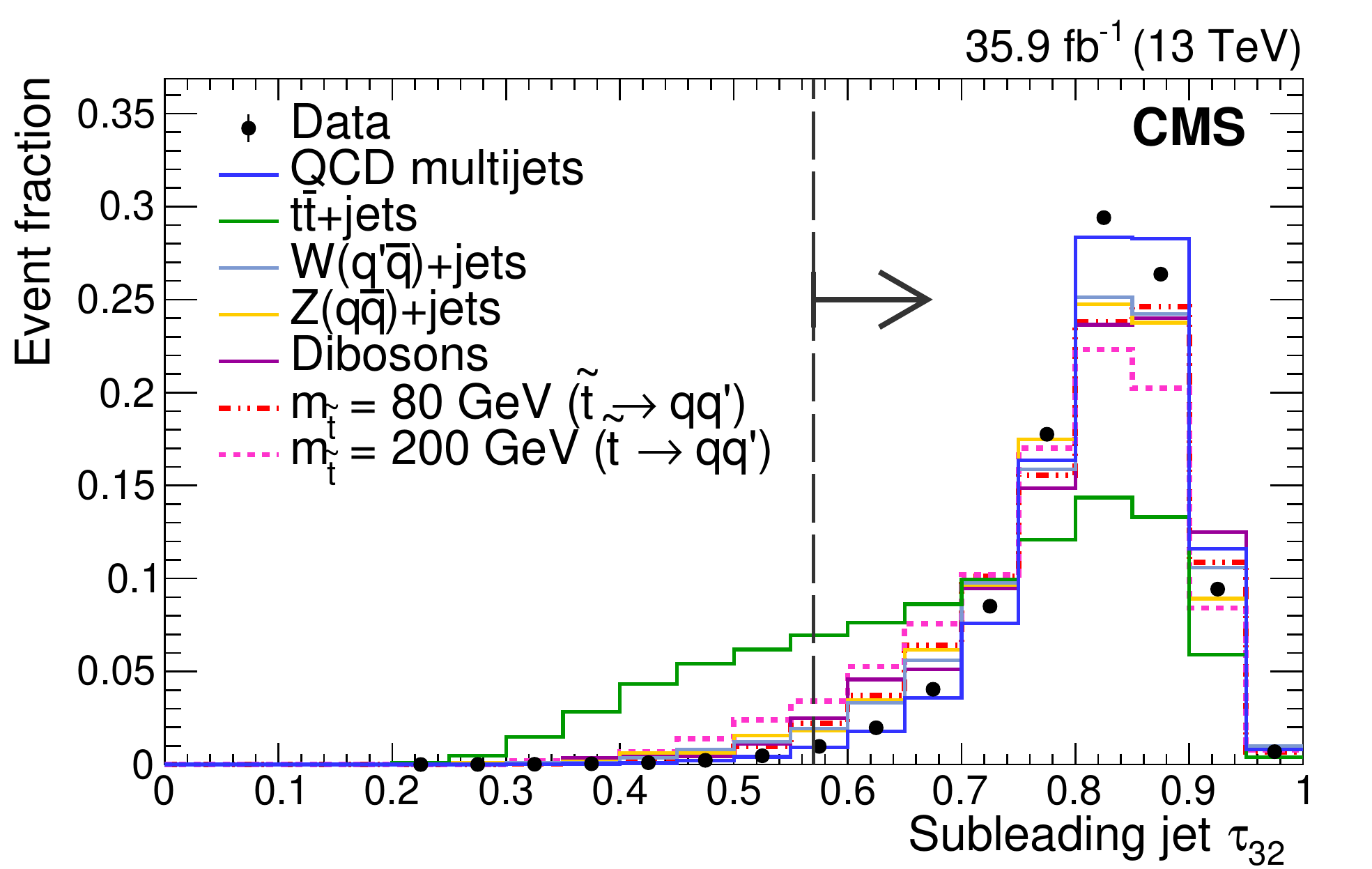}
	\caption{Boosted search kinematic distributions, normalized to unity, showing the comparison between data (black dots), backgrounds (solid colored lines), and a few selected \stopqq signal simulated samples (dashed colored lines).
	All inclusive selection criteria are applied, apart from that on the variable being presented.
	In the case of the \tauN{21} and \tauN{32} variables, both \tauN{21} and \tauN{32} requirements are removed.
	The black dashed lines indicate the maximum value imposed by the selection in the upper and middle rows of plots, and the minimum allowed value in the lower plots.
	Upper left: \masym. Upper right: \deta. Middle left: leading jet \tauN{21}. Middle right: subleading jet \tauN{21}. Lower left: leading jet \tauN{32}. Lower right: subleading jet \tauN{32}.}
\label{fig:boostedVar}
\end{figure*}

The following selection criteria are applied to reduce SM background events.
These criteria were optimized by maximizing the signal significance using $S/\sqrt{B}$ as the metric within a mass window centered at the generated \stopmass, where $S$ and $B$ are the number of signal and background events, respectively, from simulation.
The number of events with large mass imbalance between the two signal jet candidates is reduced by selecting events with mass asymmetry, defined as $ \masym = \abs{m_{\mathrm{j1}} - m_{\mathrm{j2}}}/(m_{\mathrm{j1}} + m_{\mathrm{j2}})$, below $0.1$.
Both jets are required to satisfy $\tauN{21} < 0.45$ and $\tauN{32} > 0.57$, to reject backgrounds from QCD multijets events and those from hadronically decaying top quarks, respectively.
Jets from the signal events would be predominantly produced with similar $\eta$, compared to the widely spread QCD multijet production, and thus we require events to have an absolute value of the difference in $\eta$ between the two jets: $ \deta = \abs{ \eta_{\mathrm{j1}} - \eta_{\mathrm{j2}} } < 1.5$.
For the {\cPqb}-tagged selection, both jets are required to satisfy the loose {\cPqb} tagging criteria described in Section~\ref{sec:object}.
All the selection criteria are summarized in Table~\ref{tab:selection} (second column), and are found to be optimal for the range of masses considered in this search.
The discriminating power of each of these kinematic variables is illustrated in Fig.~\ref{fig:boostedVar} where normalized distributions between data, different simulated background components, and selected simulated signal samples are presented.

\begin{table}
\centering
	\topcaption{Summary of the signal selection criteria for the boosted search (second column) and resolved search (third column). The criteria are shown for the inclusive selection and the {\cPqb}-tagged selection.}
\centering
\begin{scotch}{lcc}
Selection                    & Boosted search & Resolved search \\
& $60 < \prunedAveMass < 450\GeV$  & $\aveMass > 350\GeV$ \\
& ($80\leq\stopmass<400\GeV$) & ($\stopmass\ge400\GeV$) \\ \hline
Inclusive & AK8 jets &  AK4 jets \\
and                                      & jet $\pt> 150\GeV$      &  jet $\pt>80\GeV$ \\
{\cPqb}-tagged			     & jet $\abs{\eta}<2.5$        &  jet $\abs{\eta}<2.5$ \\
				     & Number of jets $\ge 2$        &  Number of jets $\ge 4$ \\
                                     & $\boosHT>900\GeV $          & $\resHT>900\GeV $  \\
                                     & $\masym<0.1$     & $\Masym<0.1$      \\
				     & $\tauN{21}<0.45$    & $\Deta<1.0$      \\
                                     & $\tauN{32}>0.57$    & $\Delta>200\GeV$  \\
                                     & $\deta<1.5$     &                          \\ [\cmsTabSkip]
{\cPqb}-tagged  & two loose {\cPqb}-tagged jets &  two loose {\cPqb}-tagged jets  \\
\end{scotch}
\label{tab:selection}
\end{table}

\subsection{Signal efficiency}
\begin{figure*}
\begin{center}
\includegraphics[width=0.49\textwidth]{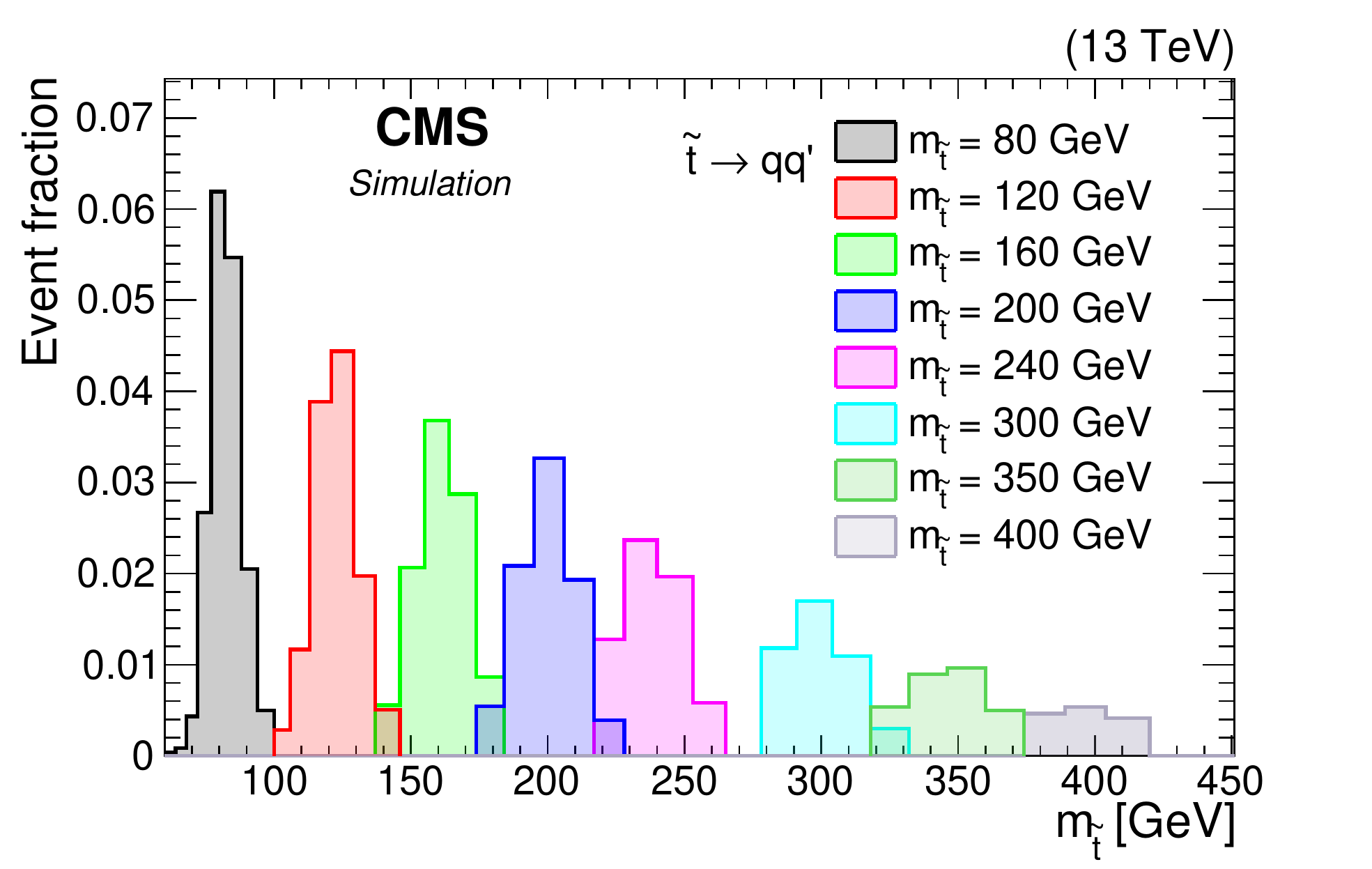}
\includegraphics[width=0.49\textwidth]{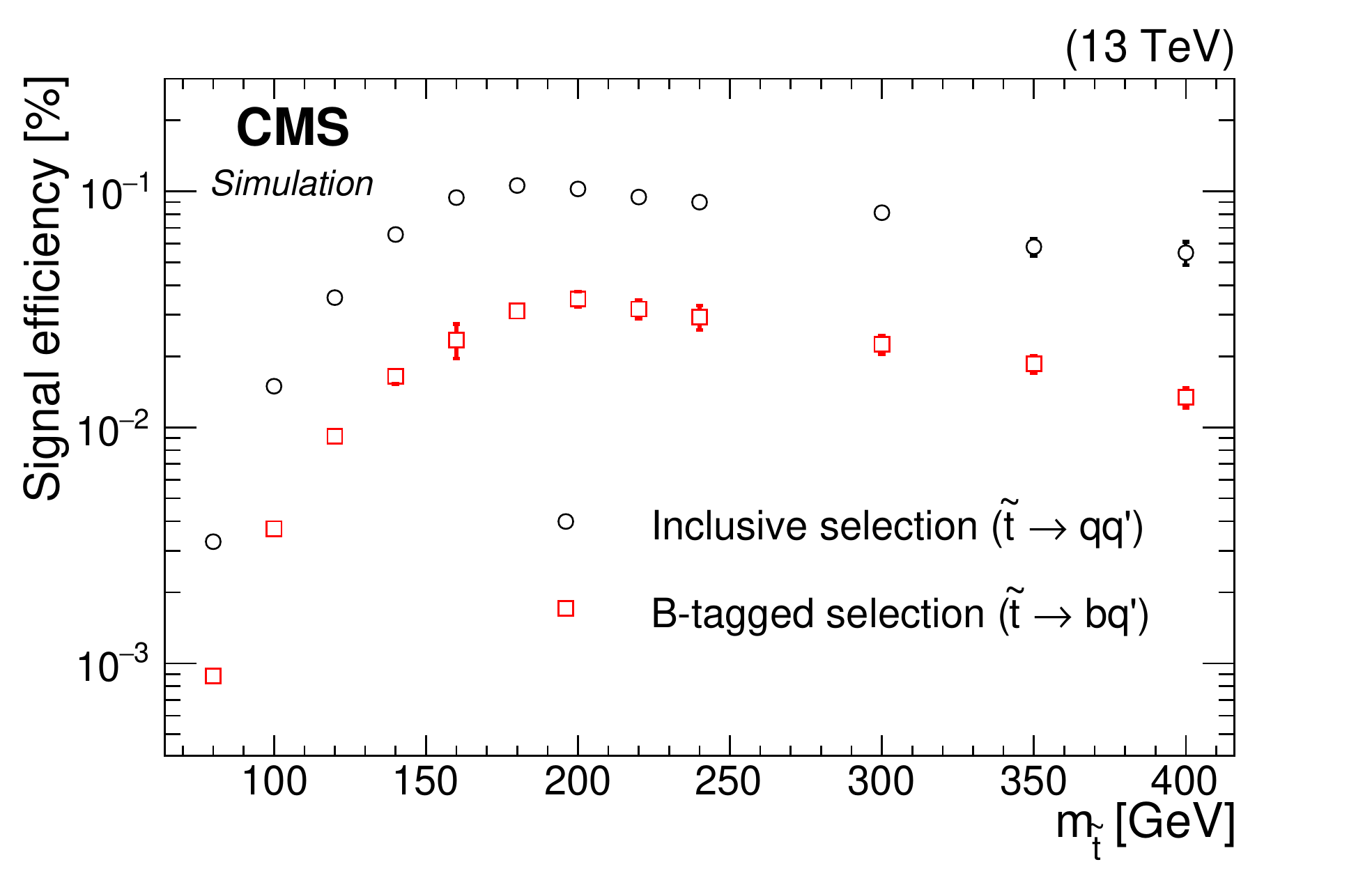}
	\caption{Boosted search signal distributions. Left: signal mass distributions after applying the inclusive selection, for various simulated \stopqq masses probed in this analysis.
	Right: signal efficiency as a function of \stopmass for the inclusive and {\cPqb}-tagged selections.}
\label{fig:boostedSignal}
\end{center}
\end{figure*}

Figure~\ref{fig:boostedSignal} (left) shows the mass distributions for simulated signals after the inclusive selection.
Similar signal mass shapes are found when applying the {\cPqb}-tagged selection.
Additionally, the signal efficiency for the boosted search is reported in Fig.~\ref{fig:boostedSignal} (right) for both the inclusive and {\cPqb}-tagged selections.
The fraction of \stopqq signal events remaining after applying the inclusive selection, relative to the total number of events generated, is 0.003\% for $\stopmass = 80\GeV$, increases to 0.106\% for $\stopmass = 180\GeV$, and drops again to 0.055\% for $\stopmass = 400\GeV$ because of the decrease in the production of top squarks with large Lorentz boosts at higher masses.
Although the fraction of boosted resonances is higher for $\stopmass \lesssim 170\GeV$, the \HT and \pt trigger requirements have a considerable impact on the event selection and are the main source of the signal efficiency loss.
The low signal selection efficiencies for boosted resonances are compensated by the large signal cross sections for low-mass top squarks~\cite{Borschensky:2014cia,nllfast31}.
The {\cPqb}-tagged selection presents a similar pattern, where the fraction of remaining events for \stopbq is 0.0009\%, 0.0350\%, and 0.0134\% for the resonance masses $\stopmass = 80$, $200$, and $400\GeV$, respectively.

\subsection{Background estimate}
After all the selection criteria are applied, the dominant remaining SM background is QCD multijet production.
Subdominant resonant backgrounds are estimated from simulation and they include {\ttbar}+jets, $\PW{\to}\qpq{+}\text{jets}$, $\cPZ{\to}\qqbar{+}\text{jets}$,
and diboson ({\PW}{\PW}, {\cPZ}{\cPZ}, {\PW}{\cPZ}) production.
The normalization of {\ttbar}+jets, the largest resonant background, is assessed in a control region enriched in {\ttbar} events by requiring $\tauN{32}<0.57$.
This criterion aims to remove one- or two-prong jets, thus enriching the sample in \ttbar.
We then compare the \prunedAveMass spectrum between data and simulation and obtain a correction factor from a first-order polynomial fit subtracting all other backgrounds.
This correction is found to be flat in \prunedAveMass and consistent with unity within 10\%, and is used as an estimate of the systematic uncertainty associated with modeling the simulated SM events.
In addition, the statistical uncertainty due to the limited number of simulated SM events in each bin is considered as a systematic uncertainty, affecting the shape of the \prunedAveMass distribution.

The background originating from QCD multijet events is estimated by extrapolating data in sideband regions to the signal region, using two uncorrelated variables and is referred to as the \textit{ABCD} method.
The variables \masym and \deta are found to have a correlation in data and in simulation of less than 1\%, therefore, these two variables are used to define four regions summarized in Table~\ref{tab:ABCDdesc}.
Region $A$ is the signal region defined by the nominal inclusive selection criteria, while the other three regions are background dominated.
Regions $B$ and $C$ are sideband regions where the event must pass one of the two selection criteria and fail the other is applied, and region $D$ is defined as the sideband region when both selection criteria fail.

\begin{table}[htbp]
\centering
\topcaption{Definition of the regions used in the QCD multijet background estimate for the boosted analysis. Region $A$ is the signal-dominated region while regions $B$, $C$, and $D$ are background-dominated sideband regions. }
\begin{scotch}{lcc}
& $\masym< 0.1$ & $\masym> 0.1$  \\ \hline
$\deta > 1.5$  & $B$ & $D$  \\
$\deta < 1.5$  & $A$ & $C$ \\
\end{scotch}
\label{tab:ABCDdesc}
\end{table}

\begin{figure}[hbtp]
\centering
\includegraphics[width=\cmsFigWidthTwo]{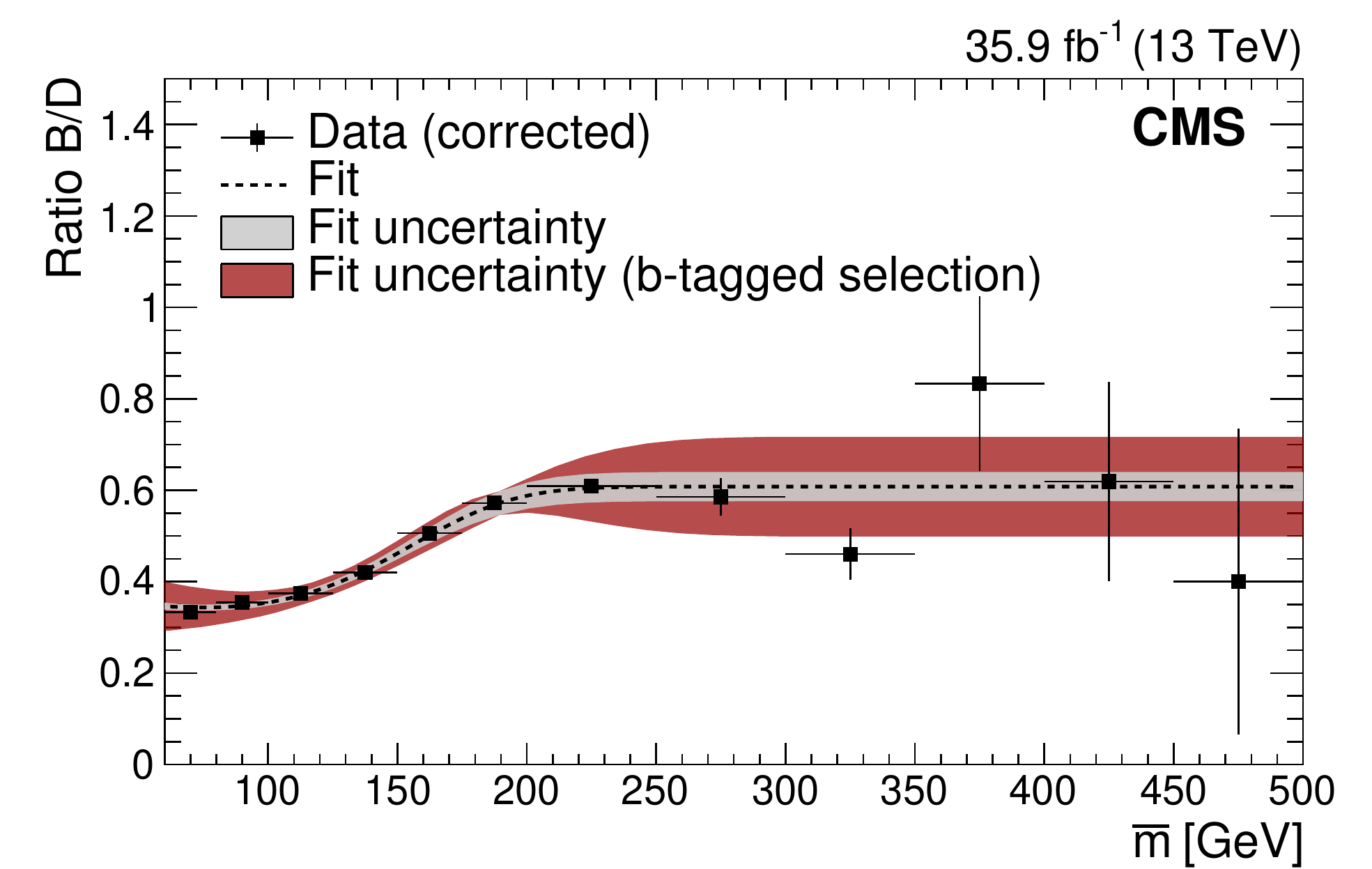}
	\caption{Boosted search transfer factor $B/D$ as a function of \prunedAveMass for data (black points) with the inclusive selection applied, and corrected for the resonant background component. The fit to the data (black dotted line) with the sigmoid function described in Eq.~\eqref{eq:tf} is also displayed. Light gray and dark red bands represent the uncertainties of the fit for the inclusive and {\cPqb}-tagged selection, respectively, and are treated as systematic uncertainties.}
\label{fig:ratioBD}
\end{figure}

The yield and the shape of the \prunedAveMass spectrum for the QCD multijet background in the signal region ($A$) is determined using the mass spectra in sideband regions such that $A = B C / D$.
The transfer factor is defined as the ratio $B/D$ and it is parameterized empirically as a function of \prunedAveMass using a sigmoid function of the form
\begin{linenomath}
\begin{equation}
	f(\prunedAveMass) = \frac{1}{p_0 + \exp( p_1 + p_2 \prunedAveMass^2 - p_3\prunedAveMass^3 )}\quad, \label{eq:tf}
\end{equation}
\end{linenomath}
where the coefficients $p_0$ to $p_3$ are free parameters of the function.
Resonant background contributions estimated from simulation are subtracted from the data prior to the extrapolation.
The fit of the transfer factor is found to give consistent results in data and simulation.
The resulting fit in the data, shown in Fig.~\ref{fig:ratioBD}, is applied to events in region $C$ to estimate the final \prunedAveMass distribution for QCD multijet events in region $A$ for the inclusive selection.
The uncertainty in the fitted transfer factor and the statistical uncertainty in the \prunedAveMass distribution in region $C$ are treated as systematic uncertainties that affect the shape of the \prunedAveMass distribution.

For the {\cPqb}-tagged selection, an equivalent procedure is performed.
Once the {\cPqb} tagging is applied, the data sample is found to be too small to obtain a transfer factor.  Instead, the transfer factor from the inclusive selection is used, and applied to region $C$ where the {\cPqb} tagging requirement is added.
By comparing the fit parameters of the transfer factors obtained with the inclusive and the {\cPqb}-tagged selections, an additional uncertainty is applied to cover the differences, as illustrated in the dark red band of Fig.~\ref{fig:ratioBD}.

\subsection{Systematic uncertainties}
The performance of the \textit{ABCD} background estimate is tested on simulated QCD multijet events.
In this test, the background prediction is compared to the mass spectrum in the signal region $A$.
The level of agreement between these two distributions, or closure, is found to be within $\pm$10\% over the entire \prunedAveMass spectrum.
This is used as an estimate of the contribution from this source to the systematic uncertainty in the QCD multijet background for both the inclusive and {\cPqb}-tagged selection.

The systematic uncertainties in the background estimates are summarized in Table~\ref{tab:bkgsyst}.

\begin{table*}
\centering
\topcaption{Summary of the sources of systematic uncertainties in the background predictions. The values of the systematic uncertainties and whether they affect the overall event yield or the shape of the mass spectra are shown.  For uncertainties affecting the shape, the range of values quoted represent the minimum and maximum effects over all bins. For the QCD multijets fit uncertainty of the resolved search, the values quoted are the combined effects of the shape and yield uncertainties of this background estimate. The symbols $^\ast$ and $^\dagger$ denote uncertainties specific to the inclusive and  {\cPqb}-tagged selections, respectively.  }
\begin{scotch}{lclcc}
Search			& Background		& Source of systematic uncertainty			& Effect &  Value         \\ \hline
Boosted & QCD & Closure 				& Yield 	& 10.0\%  \\
	& multijets & Transfer factor fit & Shape 	& 1.0--4.0\%$^\ast$  \\
	& & &                                       	& 3.0--8.0\%$^\dagger$  \\
	& & Event count in region $C$ 	& Shape 	& 1.0--23.0\%$^\ast$   \\
	& & &                                          	& 2.0--33.0\%$^\dagger$   \\ [\cmsTabSkip]
	& Resonant & Simulation modeling & Yield 	& 10.0\% \\
	& & Statistical precision of simulation		& Shape 	& 1.0--30.0\%$^\ast$  \\
	& & &                                   	& 8.0--57.0\%$^\dagger$ \\ ~\\ [\cmsTabSkip]
\multirow{1}{*}{Resolved } & QCD & Fit parameters & Shape & 3.0--28.0\%$^\ast$ \\
 & multijets & & and Yield &  2.0--38.0\%$^\dagger$  \\
\end{scotch}
\label{tab:bkgsyst}
\end{table*}

Systematic uncertainties affecting the expected signal yield arise from the integrated luminosity measurement (2.5\%)~\cite{CMS-PAS-LUM-17-001}, the trigger efficiency (3.0\%), the modeling of the pileup interactions (1.0\%), the effect from the uncertainties in the PDF (1.0\%)~\cite{Rojo:2016ymp}, and the measurement of the jet energy scale (1.2\%) and jet resolution (1.8\%)~\cite{1748-0221-6-11-P11002,Khachatryan:2016kdb}.
For the {\cPqb}-tagged selection, the uncertainty in the efficiency for identifying bottom quarks (1.0\%) contributes to the overall uncertainty in the expected signal yield~\cite{BTV-16-002}.

Systematic uncertainties due to the jet energy scale and resolution measurements also affect the shape of the \prunedAveMass spectrum (independent of the yield).
These uncertainties are determined using the reconstructed jet mass in hadronically decaying boosted {\PW} bosons,  where differences in scale (2.0\%) and resolution (14.0\%) between data and simulation have previously been observed~\cite{Sirunyan:2017acf}. We take these differences as estimates of the associated systematic errors.

Previous studies~\cite{Khachatryan:2014vla,Sirunyan:2017acf} have shown disagreement in the pruned jet mass spectra between data and simulation when a \tauN{21} requirement is applied.
The method used to quantify this discrepancy is described in Ref.~\cite{Khachatryan:2014vla}, and is based on measuring the efficiency of identifying boosted two-prong {\PW} bosons in semileptonic \ttbar events.
For $\tauN{21}<0.45$, the ratio of the efficiencies in data and simulation, or scale factor, is measured to be $1.10 \pm 0.13$.
Since this search requires two jets to satisfy the same \tauN{21} selection, the square of the scale factor is applied to the signal events in simulation, resulting in a total two-prong scale factor of $1.21 \pm 0.29$.
A similar effect has been reported when applying the \tauN{32} requirement~\cite{CMS-PAS-JME-16-003}.
In this case, a tag-and-probe procedure is used to measure the efficiency of identifying boosted three-prong hadronic top quarks in semileptonic \ttbar events.
For $\tauN{32}<0.54$, the ratio of the efficiencies in data and simulation is $1.07\pm0.05$, and the efficiency for selecting misidentified boosted top quarks is 20\%.
However, in this search, we veto three-prong jets by requiring $\tauN{32}>0.54$, which results in an anti-three-prong scale factor of $0.99\pm0.01$ for one jet, and $0.96\pm0.02$ when two jets satisfy this \tauN{32} requirement.
The uncertainties in the two-prong (\tauN{21}) and the anti-three-prong (\tauN{32}) scale factors are propagated as systematic uncertainty in the signal yield.

Finally, the uncertainties due to the limited numbers of simulated signal events also contribute to the systematic uncertainty affecting the shape of the \prunedAveMass distribution.
A summary of the systematic uncertainties affecting the signal yield and shape are summarized in Table~\ref{tab:signalsyst}.

\begin{table*}
\centering
\topcaption{Summary of the sources of systematic uncertainties for the signal samples. The values of the systematic uncertainties and whether they affect the overall event yield or the shape of the mass spectra are shown.  For uncertainties affecting the shape, the range of values quoted represent the minimum and maximum effects over all bins. The symbols $^\ast$ and $^\dagger$ denote uncertainties specific to the inclusive and  {\cPqb}-tagged selections, respectively.}
\begin{scotch}{l l  c  r}
Search	& Source of systematic uncertainty		& Effect		& Value         \\ \hline
Boosted & Integrated luminosity  		& Yield        		& 2.5\%      \\
and	& Trigger 			& Yield 		& 3.0\% \\
resolved         	& Pileup  			& Yield       		& 1.0\%  \\
	& PDF 				& Yield 		& 1.0\% \\
	& Jet energy scale 		& Yield 		& 1.2--1.5\% \\
	&                 		& Shape 		& 2.0\%\\
       	& Jet energy resolution 		& Yield 		& 1.8--6.0\% \\
       	&                  		& Shape 		& 10.0--14.0\%  \\
	& Statistical precision of simulation 			& Shape  			&  3.0--37.0\%$^\ast$  \\
	& &                                                             			&  6.0--55.0\%$^\dagger$  \\
	& {\cPqb} tagging efficiency   	& Yield 	& 1.0\%$^\dagger$      \\  ~\\ [\cmsTabSkip]
Boosted 	& Two-prong scale factor & Yield 		& 23.0\% \\
	& Anti-three-prong scale factor & Yield 		& 2.0\% \\
\end{scotch}
\label{tab:signalsyst}
\end{table*}

\begin{figure*}[hbtp]
\centering
\includegraphics[width=\cmsFigWidthTwo]{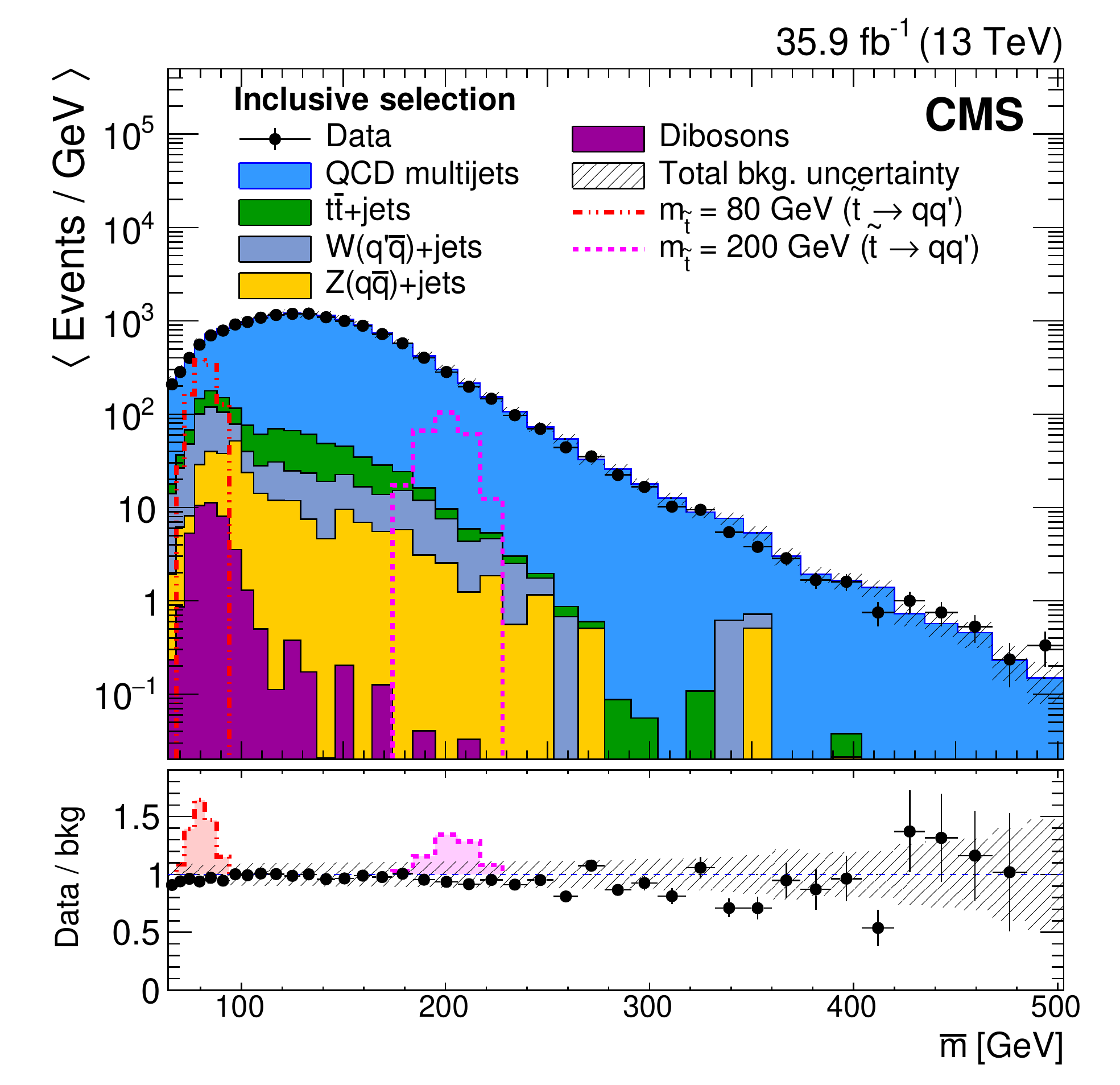}
\includegraphics[width=\cmsFigWidthTwo]{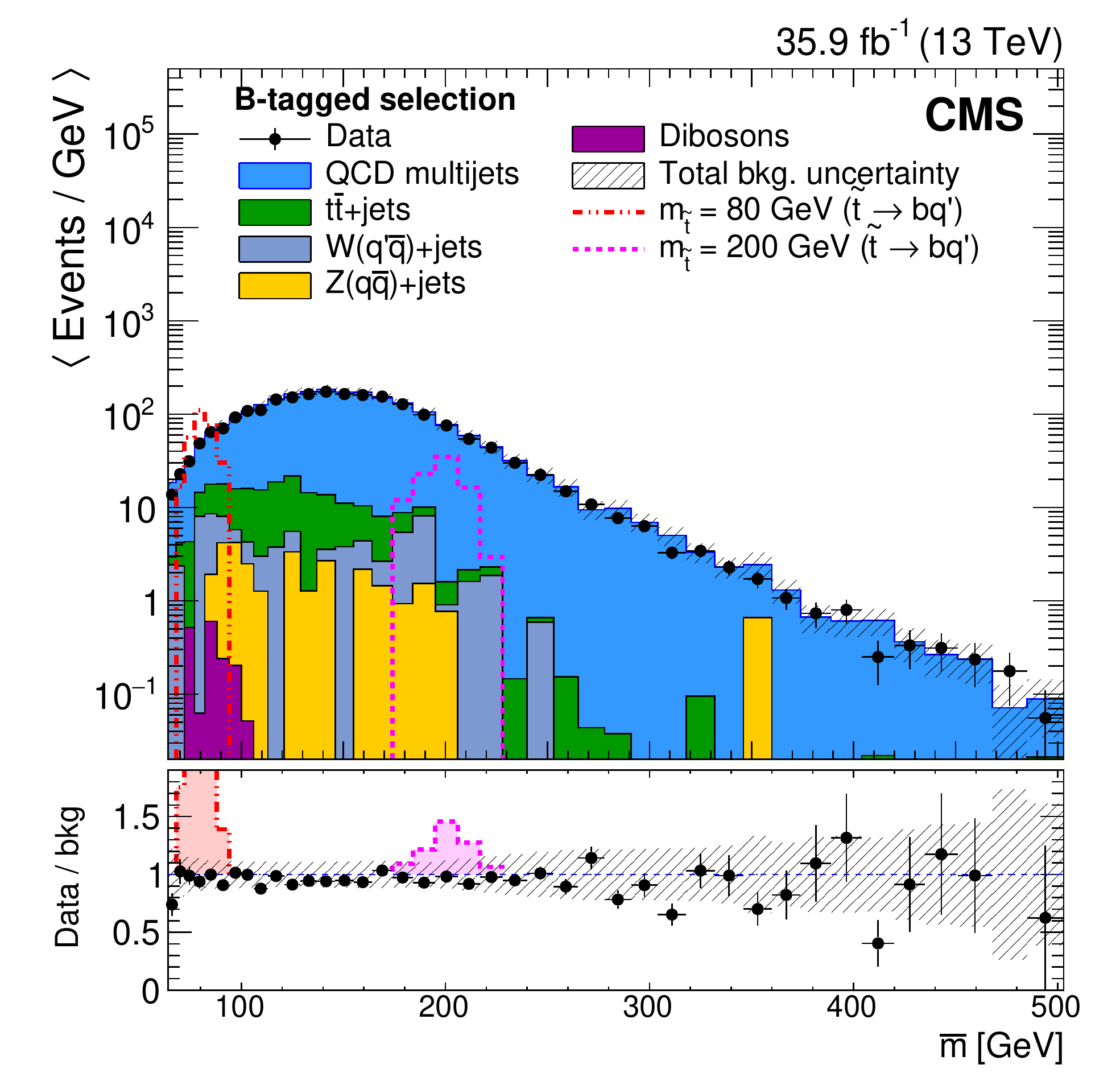}
	\caption{Boosted search \prunedAveMass distribution for data (black points) and for the total background prediction, for the inclusive (left) and the {\cPqb}-tagged (right) selection.
	The different background components are presented with different colors, while the grey hashed band displays the total background uncertainty.
  The expected signals from simulated \stopqq and \stopbq samples at $\stopmass =80\GeV$ and $\stopmass=200\GeV$  are  also  displayed (shaded  lines) for the inclusive  selection and  the  {\cPqb}-tagged selections, respectively.
	The lower panel shows the ratio between data and the background prediction.
  The shaded peaks in the lower distributions show the expected effect produced by the presence of a top squark signal, for two different top squark masses.
}
\label{fig:finalBoosted}
\end{figure*}

Figure~\ref{fig:finalBoosted} illustrates the average pruned jet mass spectrum for data and the background predictions for the inclusive (left) and the {\cPqb}-tagged (right) selections.
The resonant backgrounds correspond to less than 8\% of the total background prediction for the inclusive category, and less than 6\% for the {\cPqb}-tagged one, over the entire mass range. The data are found to agree with SM expectations.

\section{Resolved search}

\subsection{Event selection}
Events are selected using a logical OR of the \resHT trigger, described in Section~\ref{sec:boosted}, and two additional triggers: one requiring at least four AK4 jets with $\pt>50\GeV$, $\abs{\eta}< 2.5$, and $\resHT>800$\GeV, and another requiring at least four jets with $\pt>70$\GeV, $\abs{\eta}< 2.5$, and $\resHT>750$\GeV.
In addition to satisfying the trigger conditions, selected events are required to have at least four AK4 jets with $\pt> 80\GeV$, $\abs{\eta}<2.5$, and $\resHT>900\GeV$.
The selection efficiency of the chosen triggers is determined relative to unbiased data samples selected with muon based triggers.
The trigger efficiency for events that would satisfy the subsequent selection is measured to be greater than 98\%.

In order to select the two best dijet systems compatible with the signal, the four leading jets ordered in \pt are combined to create three unique combinations of dijet pairs per event.
Out of the three combinations, the dijet configuration with the smallest \DeltaR is chosen.
This variable is defined as: $\DeltaR = \sum_{\mathrm{i}=1,2}\abs{\Delta R^{\mathrm{i}} - 0.8}$,  where $\Delta R^{\mathrm{i}}$ represents the separation between two jets in the $\mathrm{i}^{\mathrm{th}}$ dijet pair, $\Delta R = \sqrt{\smash[b]{(\Delta\eta)^2 + (\Delta\phi)^2}}$, and $\Delta\eta$ and $\Delta\phi$ are the differences in $\eta$ and azimuthal angle $\phi$ (in radians) between the two jets under consideration.   This variable exploits the expectation that the decay products of the signal resonance will be closer together compared to particles from uncorrelated jets.
An offset of 0.8 has been chosen in the definition of \DeltaR to avoid overlaps between jets in the dijet systems, and to minimize the selection of dijet systems composed of jets from radiated gluons.

\begin{figure}[hbtp]
\centering
\includegraphics[width=\cmsFigWidthTwo]{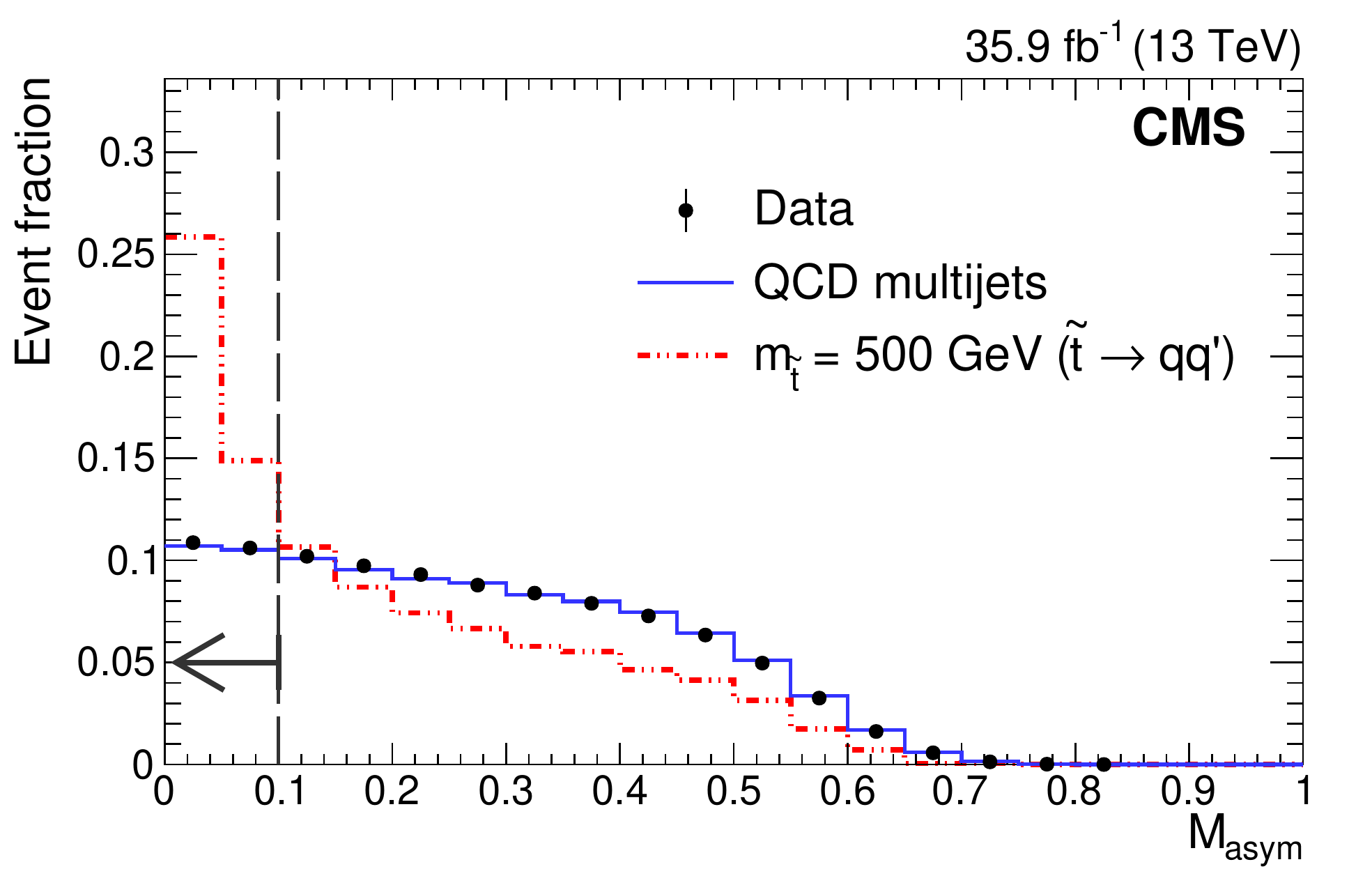}
\caption{Resolved search \Masym distribution normalized to unity for data (black dots), background (solid blue line), and a selected signal \stopqq with $\stopmass = 500\GeV$ (dashed red line).  All inclusive selection criteria are applied apart from that on the variable being presented. The region to the left of the black dashed line indicates the optimized region of selected \Masym values.
	}
\label{fig:resolvedSelmasym}
\end{figure}

\begin{figure}[hbtp]
\centering
\includegraphics[width=0.49\textwidth]{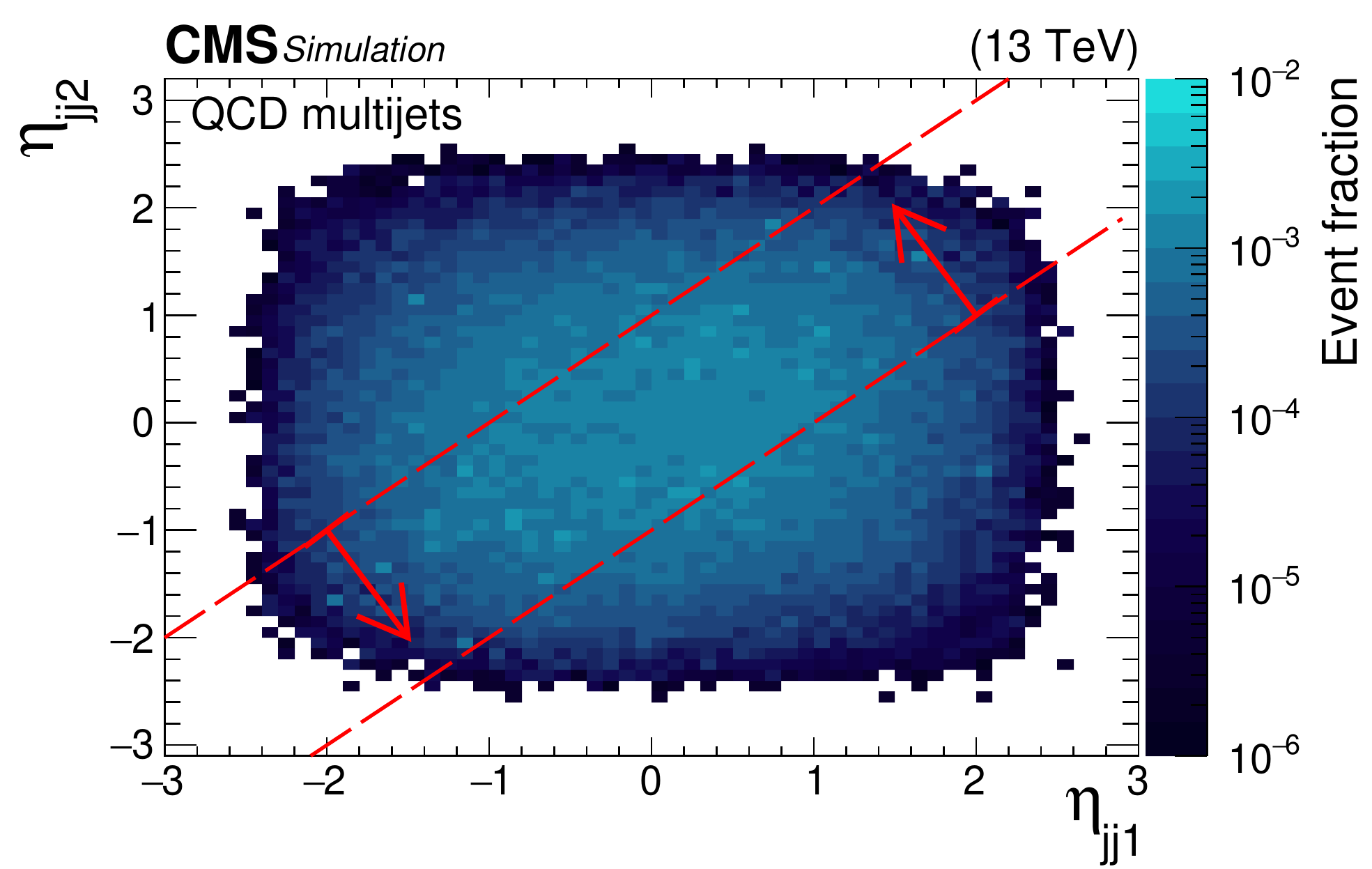}
\includegraphics[width=0.49\textwidth]{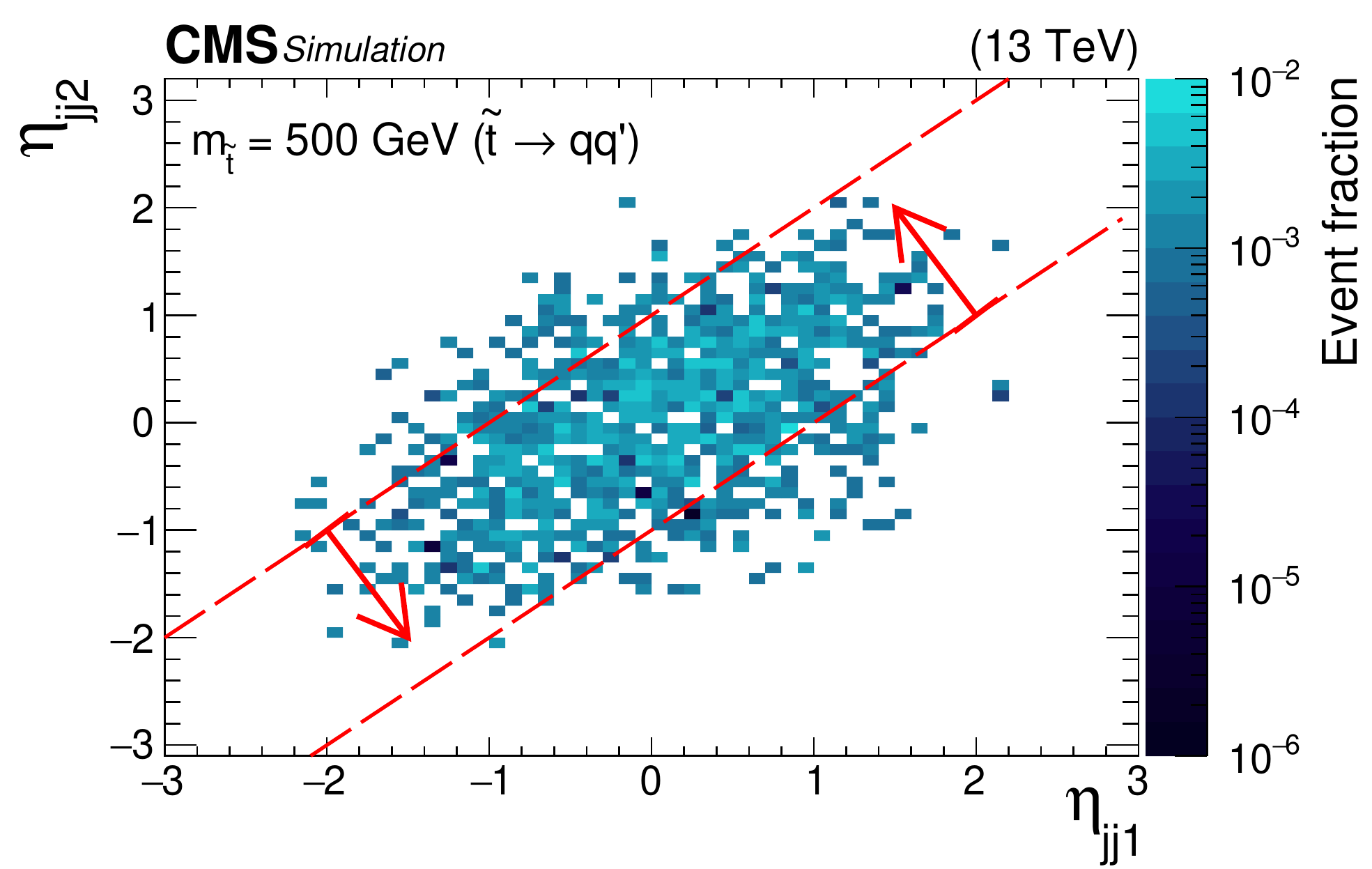}

\caption{Resolved search distribution of $\eta_{\mathrm{jj2}}$ of the lower-\pt dijet system in the selected pair as a function of the $\eta_{\mathrm{jj1}}$ of the higher-\pt dijet system.
        The distribution is shown for simulated QCD multijet events (\cmsLeft) and a representative signal \stopqq with $\stopmass = 500\GeV$ (\cmsRight).
	All inclusive selection criteria are applied apart from that on the variable being presented. The region between the two red dashed lines indicates the optimized region of selected $\Deta$ values.
	}
\label{fig:resolvedSeldeta}
\end{figure}

Once a configuration is selected, the average mass of the dijet system, $\aveMass = (\Mjj{1}+\Mjj{2})/2$, is used to search for new resonances, where $\Mjj{i}$ is the dijet mass of the $\mathrm{i}^{\mathrm{th}}$ dijet.
To further reject backgrounds from QCD multijet events and incorrect pairings from signal events, two additional requirements are applied.
As was described in Section~\ref{sec:boosted},  the dijet systems in signal events are expected to be more centrally produced than those in QCD multijet events, therefore, the pseudorapidity difference between the two dijet systems is required to be $\Deta = \abs{\eta_{\mathrm{jj1}} - \eta_{\mathrm{jj2}}} < 1.0$.
In addition, further discrimination is achieved by requiring the mass asymmetry (\Masym) between the dijet pairs to be $<0.1$, where $\Masym = \abs{ \Mjj{1} - \Mjj{2} }/( \Mjj{1} + \Mjj{2})$.
Figures~\ref{fig:resolvedSelmasym} and~\ref{fig:resolvedSeldeta} show the discriminating power of these two kinematic variables applied to data, QCD multijet simulation, and a selected simulated signal sample.

\begin{figure}[hbtp]
\centering
\includegraphics[width=0.49\textwidth]{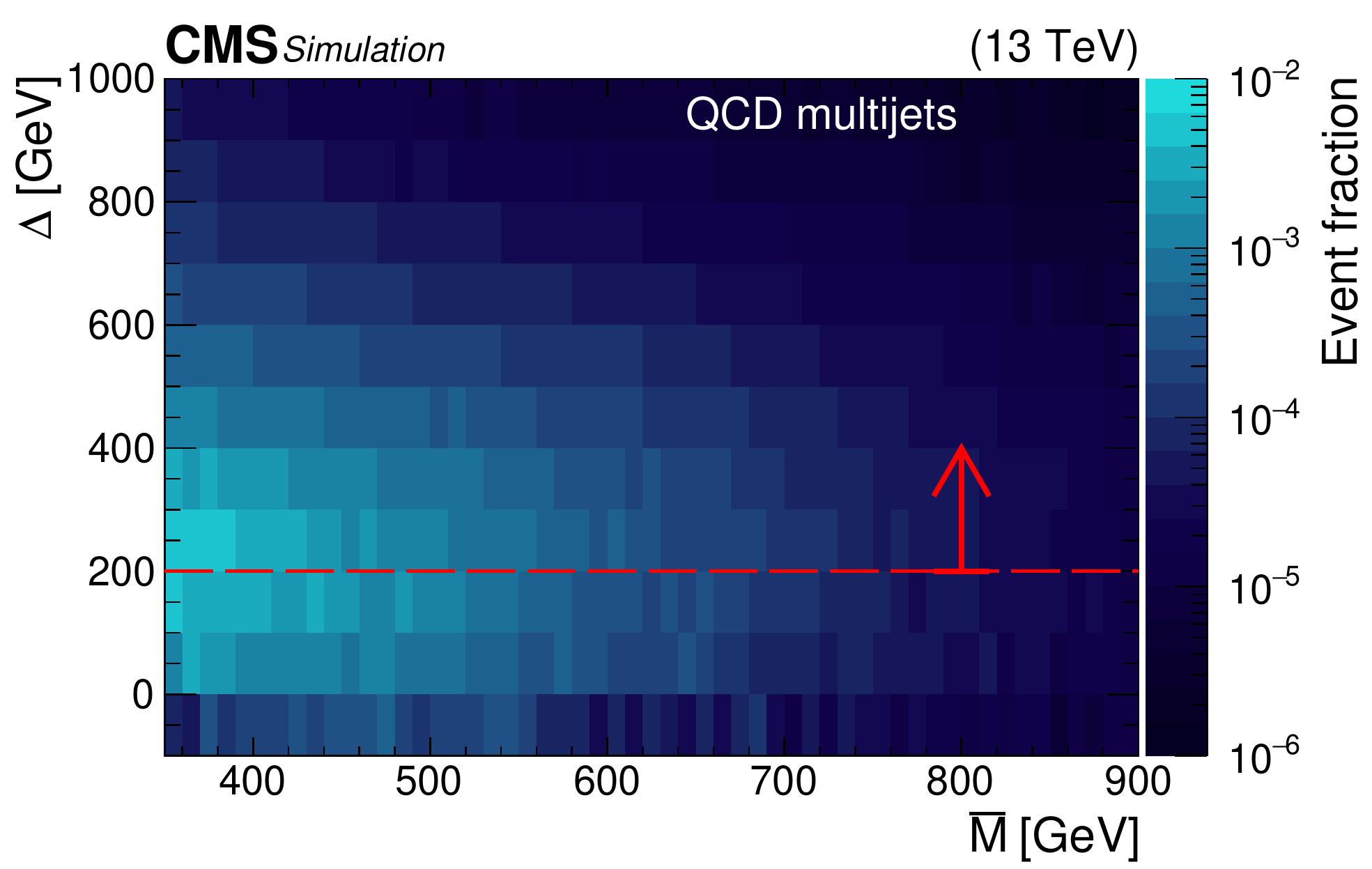}
\includegraphics[width=0.49\textwidth]{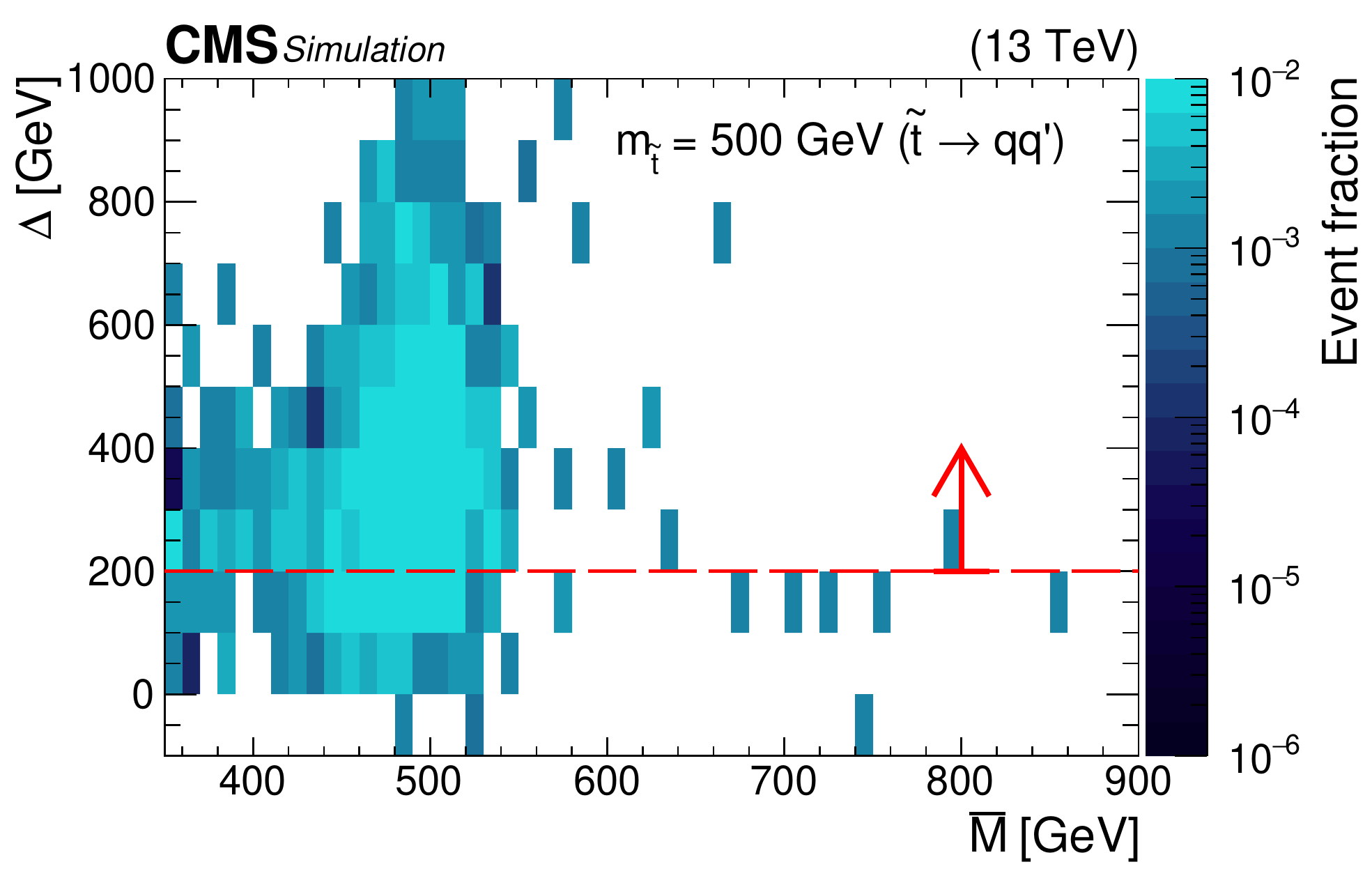}

\caption{Resolved search distribution of $\Delta$ as a function of \aveMass, shown for simulated QCD multijet events (\cmsLeft) and a representative signal \stopqq with $\stopmass = 500\GeV$ (\cmsRight).
	All inclusive selection criteria are applied apart from that on the variable being presented. The region above the red dashed line indicates the optimized region of selected $\Delta$ values.
	}
\label{fig:resolvedSeldelta}
\end{figure}

An additional variable defined as $\Delta= \bigl(\sum_{\mathrm{i}=1,2}\abs{\pt^{\mathrm{i}}}\bigr) - \aveMass$ is calculated for each dijet system, where the \pt sum is over the two jets in the dijet configuration.
The distributions of the $\Delta$ variable as a function of \aveMass for a selected signal sample and QCD multijet simulation are illustrated in Fig.~\ref{fig:resolvedSeldelta}.
This variable has been previously used in hadronic resonance searches at both the Tevatron and the LHC~\cite{cdfmultijets,cmsmultijets,cmsmultijets2,gluino2014,CMSRunIPaper7,CMSRunIPaper}.
In addition to rejecting background events, setting a minimum value of $\Delta$ results in a lowering of the peak position of the \aveMass distribution in SM QCD multijet events, and allows the search to be extend to lower resonance masses.  Events are selected with $\Delta>200\GeV$.
Finally, for the {\cPqb}-tagged selection, a loose {\cPqb}-tagged jet is required in each dijet pair candidate.
The selection requirements for this search are summarized in Table~\ref{tab:selection} (third column), and are found to be optimal for the entire range of masses considered here.

\subsection{Background estimate}
Events originating from QCD multijet processes dominate the \aveMass spectrum and are modeled with the following function
\begin{linenomath}
\begin{equation}
	\frac{{\rd}N}{{\rd}\aveMass} = \frac{p_0 (1 - x)^{p_1}}{x^{p_2}}, \label{eq:P4}
\end{equation}
\end{linenomath}
where $x=\aveMass/ \sqrt{s}$, $\sqrt{s}$ is the center-of-mass energy, $N$ is the number of considered events, and $p_0$ through $p_2$ are parameters of the function.
The functional form in Eq.~\eqref{eq:P4} successfully models the steeply falling dijet mass distribution of QCD multijet production, and comparable functions have been extensively used in similar previous dijet resonance searches~\cite{Khachatryan:2015dcf,Sirunyan:2017acf,CMSRunIPaper}.

\begin{figure*}[hbtp]
\includegraphics[width=\cmsFigWidthTwo]{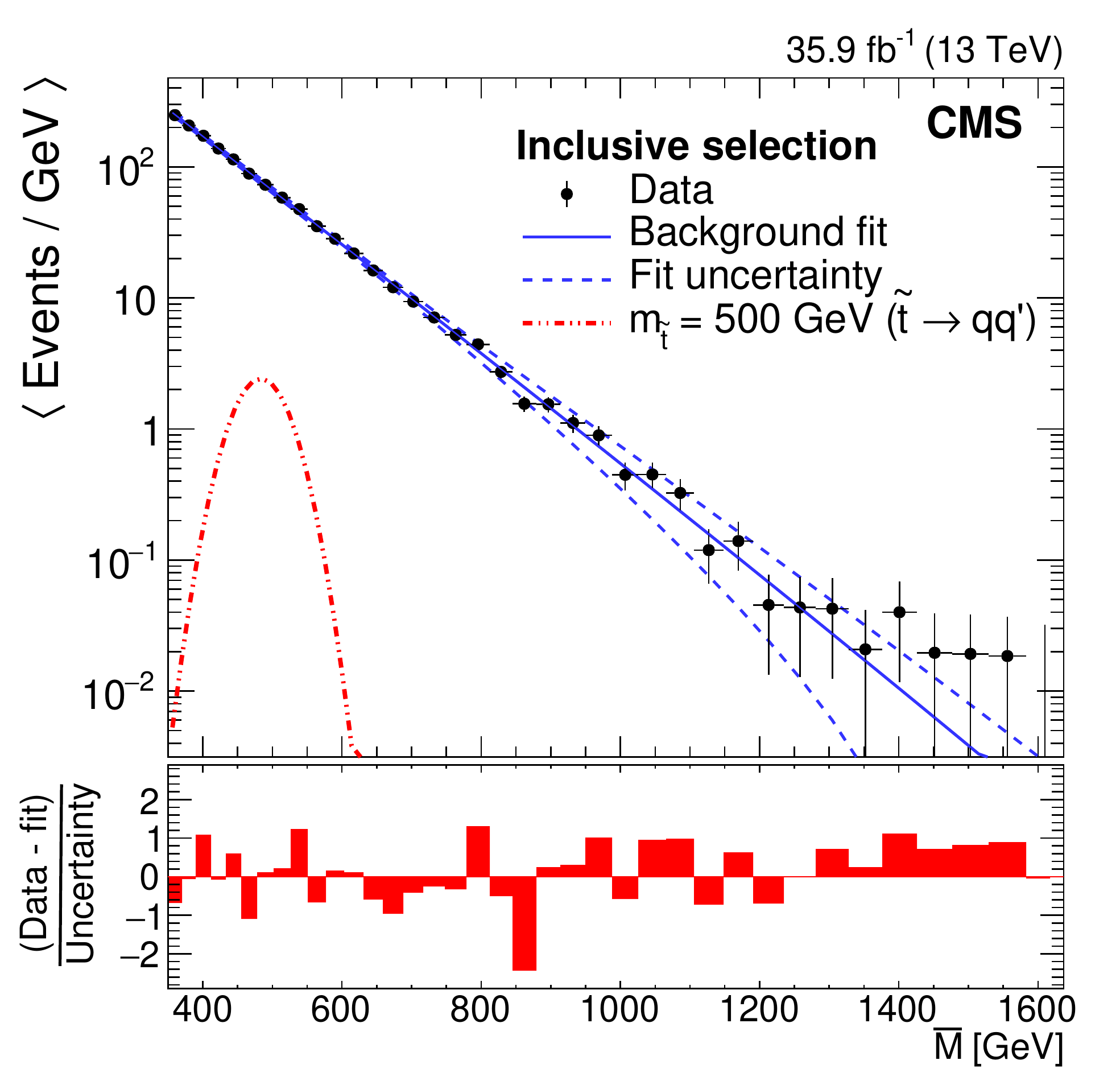}
\includegraphics[width=\cmsFigWidthTwo]{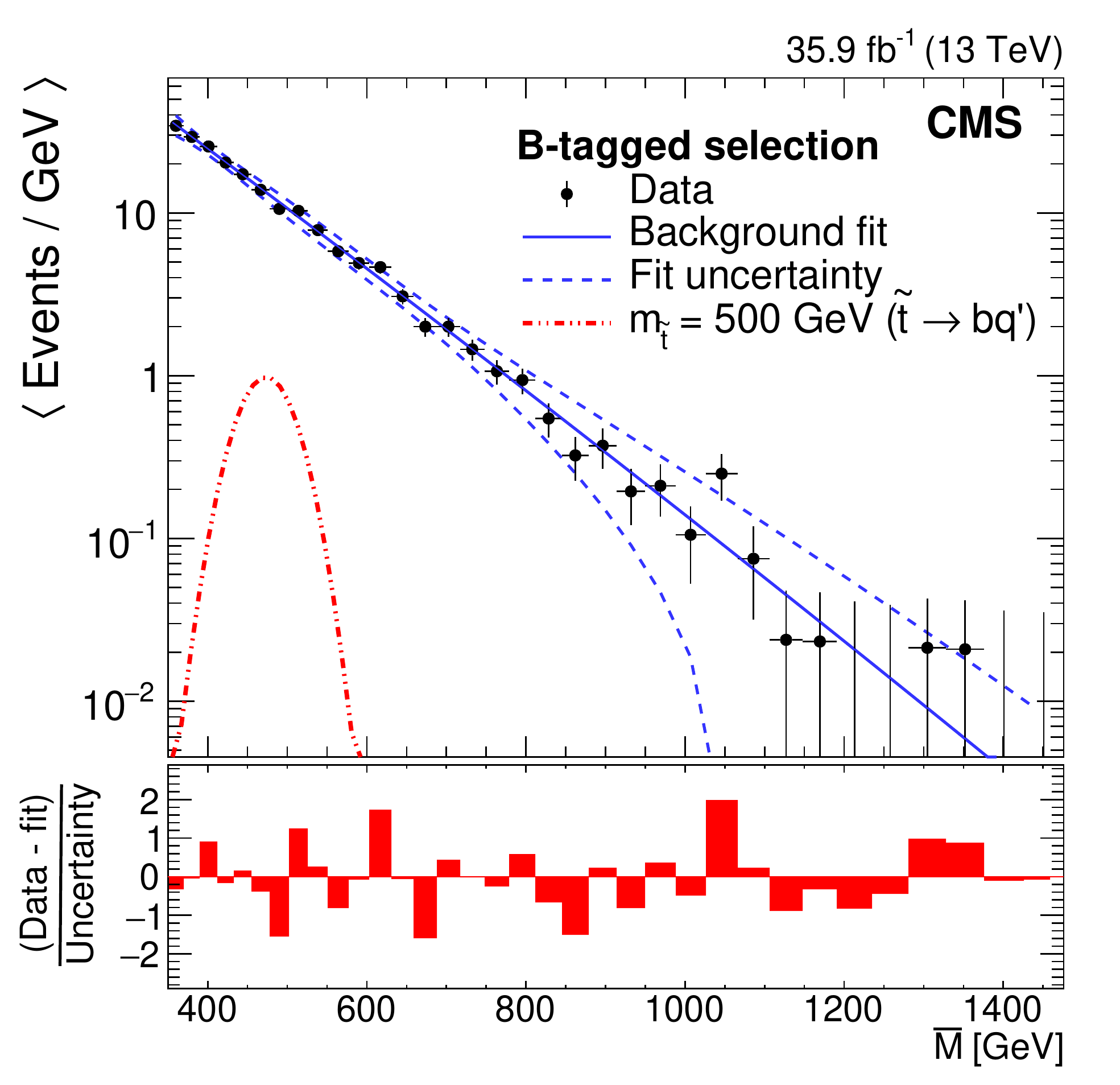}
	\caption{Resolved search distribution of \aveMass for the data (black points), along with the resulting fit to the functional form in Eq.~\eqref{eq:P4} (blue solid line) for the inclusive selection (left) and the {\cPqb}-tagged (right) selections. The expected signals from simulated \stopqq and \stopbq samples at $\stopmass = 500\GeV$ are also displayed (red dot-dashed lines) for the inclusive selection and the {\cPqb}-tagged selections, respectively. The lower panel displays the bin-by-bin difference between the data and the fit divided by the statistical uncertainty.}
\label{fig:finalResolved}
\end{figure*}

Figure~\ref{fig:finalResolved} illustrates the fitted \aveMass distributions in data using the inclusive (left) and the {\cPqb}-tagged (right) selections for the resolved analysis.
The parameterized fit is performed for \aveMass $>$ 350\GeV for both selections.
In this region the background is well modelled by the parameterization and the trigger has an efficiency greater than $98\%$ as a function of \aveMass.
Figure~\ref{fig:finalResolved} (lower panels) shows the bin-by-bin difference between the data and the fit divided by the statistical uncertainty.
The data agree with SM expectations.

The potential bias introduced by the choice of the background parameterization was investigated by performing signal injection tests in pseudo-experiments.
The pseudo-experiments were generated using the mass spectra from simulated signal events fitted with a Gaussian function, added to that of the QCD multijet simulation fitted with the function of Eq.~\eqref{eq:P4}.
Each pseudo-experiment was then fitted with alternative parameterizations from different families of functions of varying orders, and the effect on the strength of the injected signal was estimated and found to be negligible.

\subsection{Signal efficiency and systematic uncertainties}
\begin{figure*}[hbtp]
\begin{center}
\includegraphics[width=\cmsFigWidthTwo]{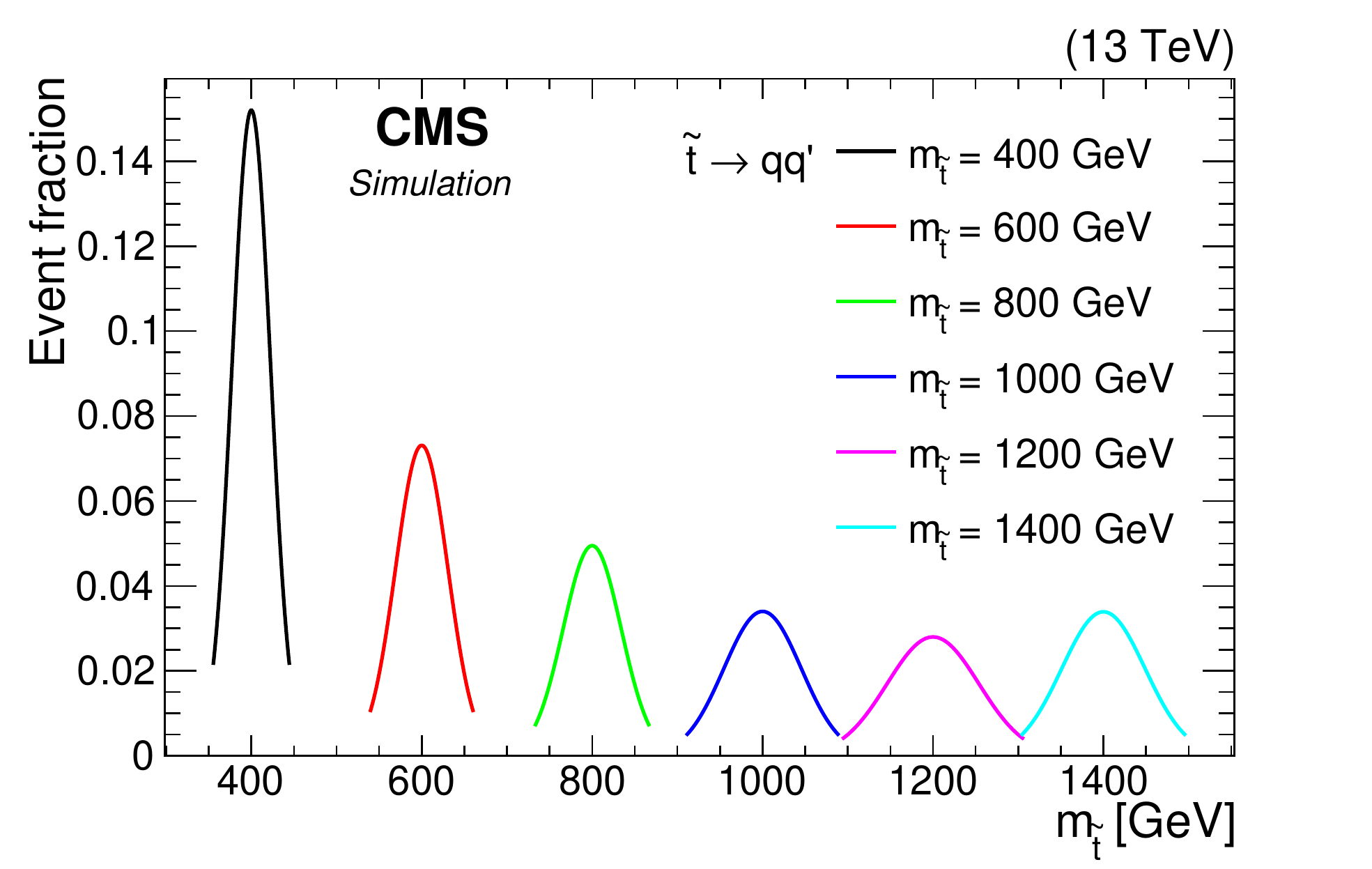}
\includegraphics[width=\cmsFigWidthTwo]{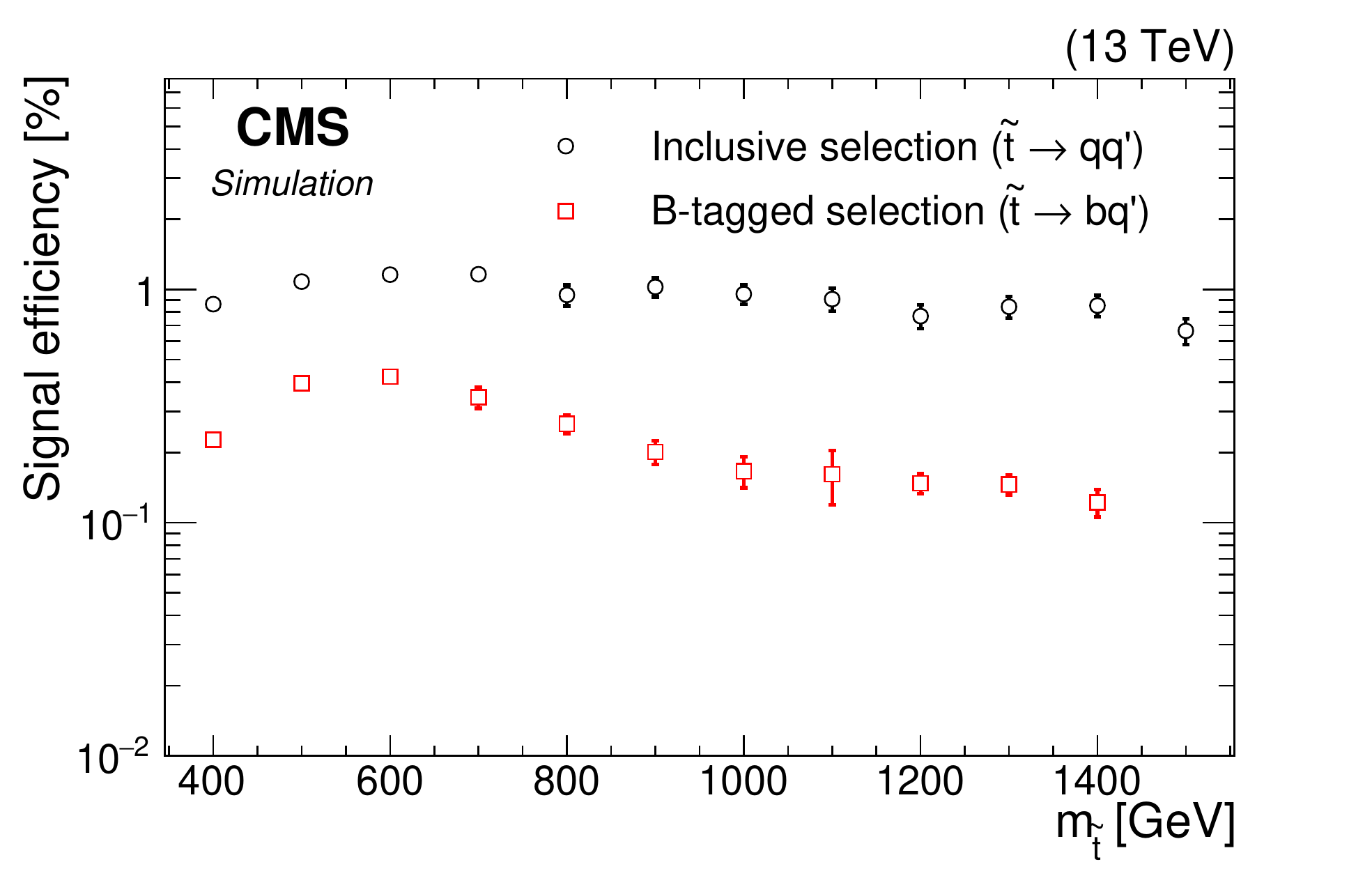}
	\caption{Simulated signal distributions for the resolved search. Left: Gaussian fits to the mass of the simulated signals for various \stopmass probed in this search for the inclusive selection.
	Right: signal efficiency, as a function of \stopmass, for the inclusive and {\cPqb}-tagged selections.}
\label{fig:resolvedSignal}
\end{center}
\end{figure*}

The \aveMass distributions of the simulated signal samples are parameterized with Gaussian functions, and are shown for the inclusive selection in Fig.~\ref{fig:resolvedSignal} (left). Similar signal mass shapes are found in the {\cPqb}-tagged analysis.
The signal efficiency for the resolved search is illustrated in Fig.~\ref{fig:resolvedSignal} (right) for both the inclusive and the {\cPqb}-tagged selections.
The fraction of \stopqq signal events remaining in simulation after applying the inclusive selection, relative to the total number of events generated, is between 0.66 and 1.16\% for \stopmass between 400 and 1500\GeV.
In the {\cPqb}-tagged selection, the fraction of remaining events in the \stopbq simulation is between 0.12 and 0.42\% for \stopmass between 400 and 1400\GeV.

The sources of systematic uncertainties affecting the normalization of the expected signal contribution are the integrated luminosity measurement (2.5\%)~\cite{CMS-PAS-LUM-17-001}, the trigger efficiency (3.0\%), the modeling of the pileup interactions (1.0\%), and the choice of PDF set (1.0\%)~\cite{Rojo:2016ymp}.   The uncertainties in the measurement of the jet energy scale (1.5\%) and resolution (6.0\%)~\cite{1748-0221-6-11-P11002,Khachatryan:2016kdb} introduce both a change in the yield and the shape of the \aveMass spectrum.
For the {\cPqb}-tagged selection, the uncertainty in the efficiency for identifying bottom quarks (1.0\%) contributes to the overall uncertainty in the expected signal yield~\cite{BTV-16-002}.
Finally, the statistical uncertainties associated with the simulated samples also contribute to the overall systematic uncertainty.
The systematic uncertainties affecting the signal are summarized in Table~\ref{tab:signalsyst}.

The uncertainties in the fitted parameters of Eq.~\eqref{eq:P4} are also taken into account as sources of systematic uncertainty affecting both the background yield and shape of the \aveMass spectrum, and are summarized in Table~\ref{tab:bkgsyst}.

\section{Results}

Figures~\ref{fig:finalBoosted} and~\ref{fig:finalResolved} present the mass spectra for the boosted and resolved analyses, respectively. They are in agreement with SM expectations.  The mass spectra are used to set limits on the pair production cross section as a function of mass of resonances decaying into quark pairs, by considering the benchmark model of top squarks decaying via the RPV couplings \UDDqq and \UDDbq. The exclusion limits are computed using the modified frequentist approach for \CL, with a binned profile likelihood as the test statistic~\cite{clsTechnique2,clsTechnique}, using an asymptotic approximation~\cite{Cowan:2010js}.

Results for the boosted search are obtained from combined signal and background binned likelihood fits to the \prunedAveMass distribution in data.
For each value of \stopmass considered, only bins of \prunedAveMass within two standard deviations of the mean of a Gaussian function fitted to the generated top squark mass are included in the likelihood.
For each bin used in the likelihood, the individual background components and the signal are allowed to float within uncertainties.
Systematic uncertainties affecting the yield and the shape, as summarized in Tables~\ref{tab:bkgsyst} and~\ref{tab:signalsyst}, are assumed to be correlated among bins.
These uncertainties are treated as nuisance parameters, which are profiled and modeled with log-normal priors, except for the uncertainty in the number of events in sideband region $C$, which is modeled with a $\Gamma$ function prior.

For the resolved search, the \aveMass spectrum in data is compared to the background fit to search for localized deviations consistent with a resonance.
For each value of \stopmass, a likelihood fit is used to compare the data to the shapes for the signal and background, within a mass window of two standard deviations around the true value of \stopmass.
Here, all systematic uncertainties are modeled with log-normal priors.

\begin{figure*}
\centering
\includegraphics[width=\cmsFigWidthTwo]{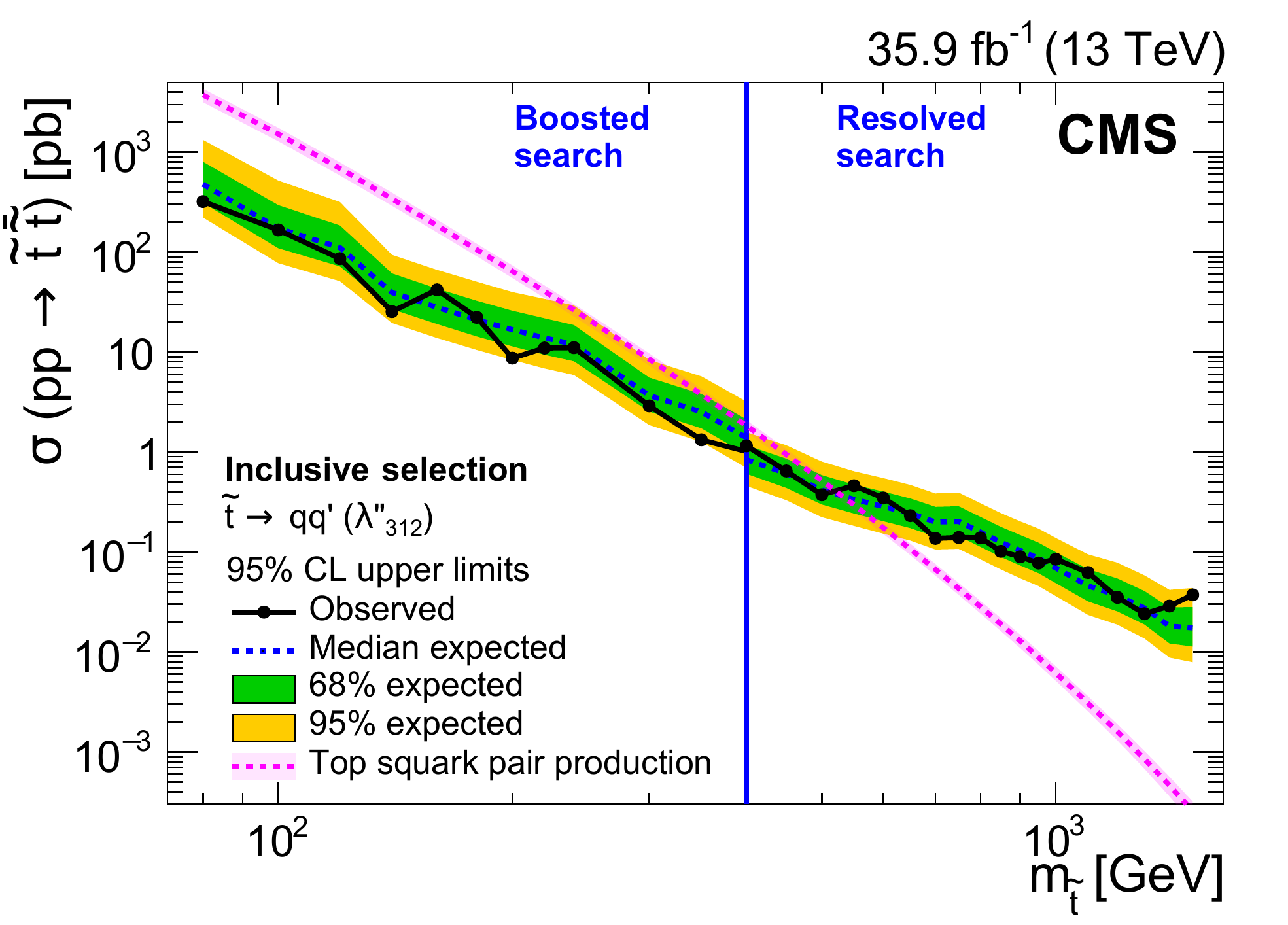}
\includegraphics[width=\cmsFigWidthTwo]{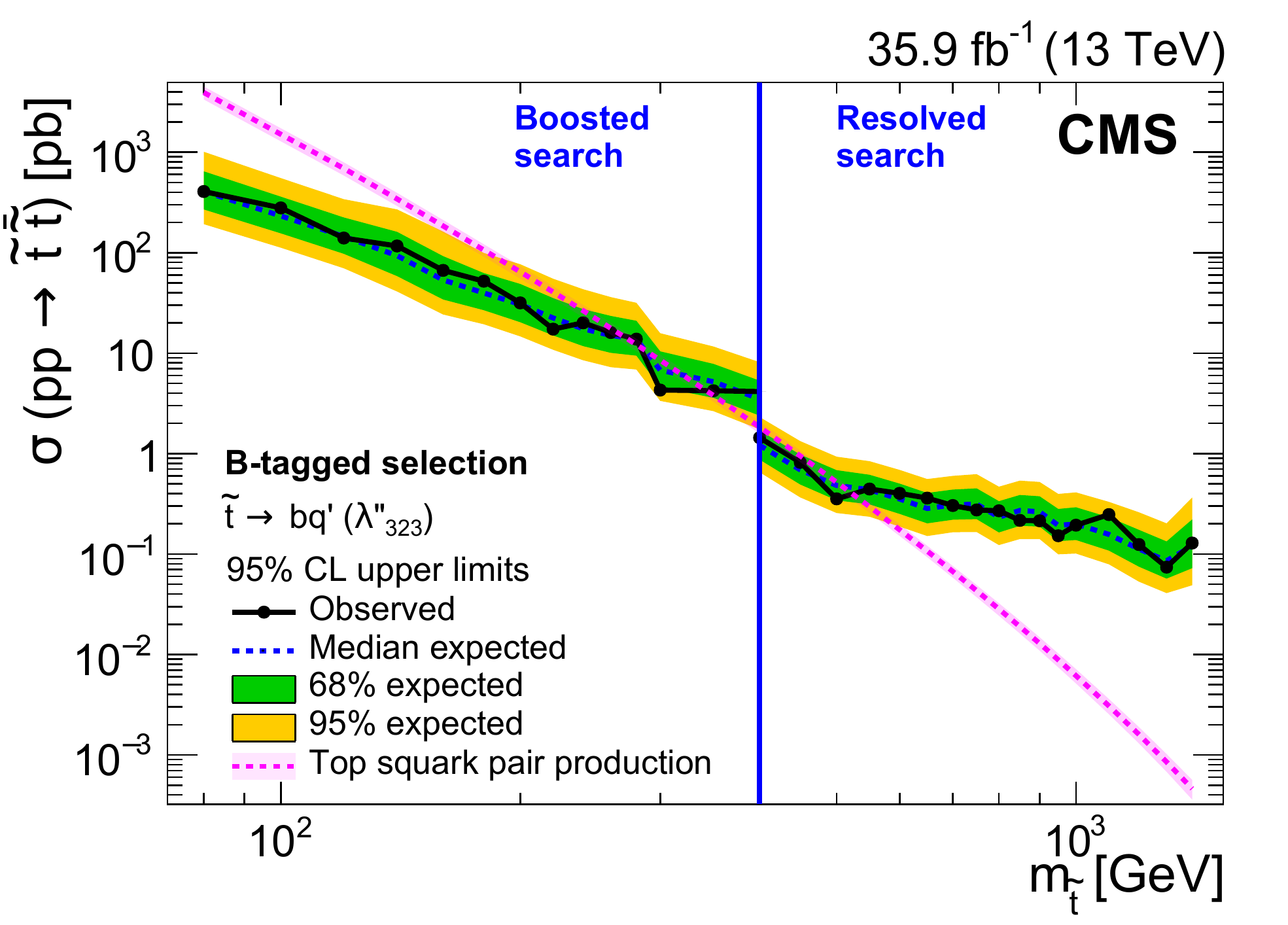}
	\caption{Observed and expected 95\% \CL upper limits on the signal cross section as a function of \stopmass. The branching fraction to quarks is assumed to be 100\%.  The boosted analysis probes $80\leq\stopmass<400\GeV$, while the resolved analysis searches for $\stopmass\ge400\GeV$. Left: limits using the inclusive selection for \stopqq assuming the RPV coupling \UDDqq. Right: limits using the {\cPqb}-tagged selection for \stopbq assuming the RPV coupling \UDDbq. The dashed pink line shows the NLO+NLL theoretical prediction for top squark pair production~\cite{Borschensky:2014cia,nllfast31}.}
\label{fig:Limits}
\end{figure*}

Figure~\ref{fig:Limits} shows the observed and expected 95\% \CL upper limits on the top squark pair production of cross section as a function of \stopmass for the boosted and resolved analyses.
The boosted analysis probes the mass range $80\leq\stopmass<400\GeV$, while the resolved analysis covers the range $\stopmass\ge400\GeV$.
Figure~\ref{fig:Limits} (left) presents the resulting limits using the inclusive selection for the \UDDqq coupling scenario, while Fig.~\ref{fig:Limits} (right) illustrates the limits using the {\cPqb}-tagged selection assuming the \UDDbq coupling.
The dashed pink line represents the theoretical prediction for the top squark pair production cross section at $\sqrt{s}=13\TeV$ evaluated at NLO with next-to-leading logarithmic (NLL) corrections~\cite{Borschensky:2014cia,nllfast31}.
We exclude top squark masses from 80 to 520\GeV assuming the \UDDqq coupling.
For the \UDDbq coupling, the boosted search excludes masses from 80 to 270 and from 285 to 340\GeV; and the resolved search excludes masses from 400 to 525\GeV.
The corresponding expected mass limits obtained are 80 to 520\GeV for top squarks decaying via \UDDqq, and 80 to 270, 285 to 320, and 400 to 505\GeV for the \UDDbq coupling.

\section{Summary}

A search has been performed for the pair production of diquark resonances in two-jet events in a boosted jet topology and in four-jet events in a resolved jet topology.
Data from proton-proton collisions at $\sqrt{s}=13\TeV$ collected in 2016 with the CMS detector, corresponding to an integrated luminosity of 35.9\fbinv, have been analysed.
In the boosted search, the distribution of the average mass of the selected two jets has been investigated for localized disagreements between data and the background estimate, consistent with the presence of a narrow resonance, while in the resolved analysis the average mass of the selected dijet pairs is utilized.
The boosted search explores resonance masses between 80 and 400\GeV, while the resolved one covers masses above 400\GeV.
We find agreement between the observation and standard model expectations.
These results are interpreted in the framework of $R$-parity-violating supersymmetry with the pair production of top squarks decaying promptly to quarks via the \UDDqq  or the \UDDbq couplings, assuming 100\% branching fractions to \stopqq or  \stopbq, respectively.
Upper limits are set at 95\% confidence level on the pair production cross section of top squarks as a function of the top squark mass.
We exclude top squark masses with the \UDDqq coupling from 80 to 520\GeV.
For the \UDDbq coupling, the boosted search excludes masses from 80 to 270 and from 285 to 340\GeV; and the resolved search excludes masses from 400 to 525\GeV.  These results probe a wider range of masses than previously explored at the LHC, and extend the top squark mass limits in the \stopqq scenario.

\begin{acknowledgments}
We congratulate our colleagues in the CERN accelerator departments for the excellent performance of the LHC and thank the technical and administrative staffs at CERN and at other CMS institutes for their contributions to the success of the CMS effort. In addition, we gratefully acknowledge the computing centers and personnel of the Worldwide LHC Computing Grid for delivering so effectively the computing infrastructure essential to our analyses. Finally, we acknowledge the enduring support for the construction and operation of the LHC and the CMS detector provided by the following funding agencies: BMBWF and FWF (Austria); FNRS and FWO (Belgium); CNPq, CAPES, FAPERJ, FAPERGS, and FAPESP (Brazil); MES (Bulgaria); CERN; CAS, MoST, and NSFC (China); COLCIENCIAS (Colombia); MSES and CSF (Croatia); RPF (Cyprus); SENESCYT (Ecuador); MoER, ERC IUT, and ERDF (Estonia); Academy of Finland, MEC, and HIP (Finland); CEA and CNRS/IN2P3 (France); BMBF, DFG, and HGF (Germany); GSRT (Greece); NKFIA (Hungary); DAE and DST (India); IPM (Iran); SFI (Ireland); INFN (Italy); MSIP and NRF (Republic of Korea); MES (Latvia); LAS (Lithuania); MOE and UM (Malaysia); BUAP, CINVESTAV, CONACYT, LNS, SEP, and UASLP-FAI (Mexico); MOS (Montenegro); MBIE (New Zealand); PAEC (Pakistan); MSHE and NSC (Poland); FCT (Portugal); JINR (Dubna); MON, RosAtom, RAS, RFBR, and NRC KI (Russia); MESTD (Serbia); SEIDI, CPAN, PCTI, and FEDER (Spain); MOSTR (Sri Lanka); Swiss Funding Agencies (Switzerland); MST (Taipei); ThEPCenter, IPST, STAR, and NSTDA (Thailand); TUBITAK and TAEK (Turkey); NASU and SFFR (Ukraine); STFC (United Kingdom); DOE and NSF (USA).

\hyphenation{Rachada-pisek} Individuals have received support from the Marie-Curie program and the European Research Council and Horizon 2020 Grant, contract No. 675440 (European Union); the Leventis Foundation; the A. P. Sloan Foundation; the Alexander von Humboldt Foundation; the Belgian Federal Science Policy Office; the Fonds pour la Formation \`a la Recherche dans l'Industrie et dans l'Agriculture (FRIA-Belgium); the Agentschap voor Innovatie door Wetenschap en Technologie (IWT-Belgium); the F.R.S.-FNRS and FWO (Belgium) under the ``Excellence of Science - EOS" - be.h project n. 30820817; the Ministry of Education, Youth and Sports (MEYS) of the Czech Republic; the Lend\"ulet (``Momentum") Program and the J\'anos Bolyai Research Scholarship of the Hungarian Academy of Sciences, the New National Excellence Program \'UNKP, the NKFIA research grants 123842, 123959, 124845, 124850 and 125105 (Hungary); the Council of Science and Industrial Research, India; the HOMING PLUS program of the Foundation for Polish Science, cofinanced from European Union, Regional Development Fund, the Mobility Plus program of the Ministry of Science and Higher Education, the National Science Center (Poland), contracts Harmonia 2014/14/M/ST2/00428, Opus 2014/13/B/ST2/02543, 2014/15/B/ST2/03998, and 2015/19/B/ST2/02861, Sonata-bis 2012/07/E/ST2/01406; the National Priorities Research Program by Qatar National Research Fund; the Programa Estatal de Fomento de la Investigaci{\'o}n Cient{\'i}fica y T{\'e}cnica de Excelencia Mar\'{\i}a de Maeztu, grant MDM-2015-0509 and the Programa Severo Ochoa del Principado de Asturias; the Thalis and Aristeia programs cofinanced by EU-ESF and the Greek NSRF; the Rachadapisek Sompot Fund for Postdoctoral Fellowship, Chulalongkorn University and the Chulalongkorn Academic into Its 2nd Century Project Advancement Project (Thailand); the Welch Foundation, contract C-1845; and the Weston Havens Foundation (USA).
\end{acknowledgments}
\bibliography{auto_generated}
\cleardoublepage \appendix\section{The CMS Collaboration \label{app:collab}}\begin{sloppypar}\hyphenpenalty=5000\widowpenalty=500\clubpenalty=5000\vskip\cmsinstskip
\textbf{Yerevan Physics Institute, Yerevan, Armenia}\\*[0pt]
A.M.~Sirunyan, A.~Tumasyan
\vskip\cmsinstskip
\textbf{Institut f\"{u}r Hochenergiephysik, Wien, Austria}\\*[0pt]
W.~Adam, F.~Ambrogi, E.~Asilar, T.~Bergauer, J.~Brandstetter, M.~Dragicevic, J.~Er\"{o}, A.~Escalante~Del~Valle, M.~Flechl, R.~Fr\"{u}hwirth\cmsAuthorMark{1}, V.M.~Ghete, J.~Hrubec, M.~Jeitler\cmsAuthorMark{1}, N.~Krammer, I.~Kr\"{a}tschmer, D.~Liko, T.~Madlener, I.~Mikulec, N.~Rad, H.~Rohringer, J.~Schieck\cmsAuthorMark{1}, R.~Sch\"{o}fbeck, M.~Spanring, D.~Spitzbart, A.~Taurok, W.~Waltenberger, J.~Wittmann, C.-E.~Wulz\cmsAuthorMark{1}, M.~Zarucki
% ...
\vskip\cmsinstskip
% ...
1:  Also at Vienna University of Technology, Vienna, Austria\\
% ...
\end{sloppypar}
\end{document}